\documentclass[aps,prl,twocolumn,groupedaddress]{revtex4}

\usepackage{graphicx}
\usepackage{docs}%
\usepackage{bm}%
\begin{document}


\title{Chaos and the Continuum Limit in Nonneutral Plasmas and Charged
Particle Beams}


\author{Henry E. Kandrup}
\email{kandrup@astro.ufl.edu}
\homepage[]{www.astro.ufl.edu/~galaxy}
\affiliation{Department of Astronomy, Department of Physics, and Institute
for Fundamental Theory, University of Florida, Gainesville, FL 32611}
\author{Ioannis V. Sideris}   
\email{sideris@nicadd.niu.edu}
\affiliation{Department of Physics, Northern Illinois University, DeKalb,
IL 60115}
\author{Courtlandt L. Bohn}
\email{clbohn@fnal.gov}
\affiliation{Department of Physics, Northern Illinois University, DeKalb,
IL 60115} 
\affiliation{Fermilab, Batavia, IL 60510}

\date{\today}

\begin{abstract}
This paper summarises an investigation of discreteness effects and the
continuum limit for time-independent, nearly collisionless $N$-body systems 
of charged particles interacting via an unscreened $1/r^{2}$ force, allowing
for bulk density distributions corresponding to potentials that admit both
regular and chaotic orbits. Both for orbit ensembles and for 
individual orbits, for $N\to\infty$ there 
is a smooth convergence towards a continuum limit. At least for moderately 
large values of $N$ discreteness effects are extremely well modeled by 
Gaussian white noise with energy relaxation time $t_{R}$, and hence diffusion 
constant $D$, consistent with the scaling 
$t_{R}{\;}{\propto}{(N/ \ln {\Lambda}})t_{D}$, with ${\Lambda}$ the Coulomb
logarithm and $t_{D}$ a natural `dynamical' time scale, as predicted by a 
Fokker-Planck description. Discreteness effects accelerate emittance growth 
for an initially localised ensemble of orbits (`clump'). 
However, even allowing for discreteness effects one can distinguish clearly
between orbits which, in the continuum limit, feel a regular (nonchaotic)
potential, so that emittance grows as a power law in time, and chaotic
orbits, for which the emittance grows exponentially. 
For sufficiently large $N$, one can implement a clear distinction between
two different `kinds' of chaos acting in $N$-body systems. Short range
{\em microchaos}, associated with close encounters between individual
charges, is a generic feature of the $N$-body problem, giving rise to 
{\em large positive Lyapunov exponents ${\chi}_{N}$ which do not decrease 
with increasing $N$ even if the bulk potential is integrable.} Alternatively, 
there is the possibility of larger scale {\em macrochaos}, characterised 
typically by somewhat smaller Lyapunov exponents ${\chi}_{S}$, which will 
be present only if the bulk potential admits global stochasticity. Conventional
computations of the largest Lyapunov exponent provide 
estimates of ${\chi}_{N}$, leading to the oxymoronic conclusion that 
$N$-body orbits which look nearly regular and have sharply peaked Fourier
spectra are `very chaotic.' However, the `range' of the microchaos
is set by the typical interparticle spacing which, as $N$ increases, becomes 
progressively smaller, so that, for sufficiently large $N$, this microchaos,
albeit very strong, is largely irrelevant macroscopically. A more 
careful numerical analysis allows one to derive estimates of both ${\chi}_{N}$ 
and ${\chi}_{S}$.
\end{abstract}


\maketitle


\section{Introduction and Motivation}
To what extent can a `nearly collisionless' $N$-body system such as a
nonneutral plasma or a charged particle beam be modeled by a smooth density
distribution and a smooth bulk potential, either statistically or at the level 
of individual orbits? 
In particular, is there a well defined $N\to\infty$ continuum limit?
The idealisation of a smooth potential is extremely convenient, both
conceptually and computationally.
However, as noted, {\em e.g.,} by beam dynamicists~\cite{Struck}, it is not
completely clear when -- or if -- such an idealisation is justified.

Even assuming that there is a well-defined continuum limit, how large must $N$
be before the continuum approximation is justified? And to what extent can
residual discreteness effects be modeled in the context of a Fokker-Planck
description? The conventional Fokker-Planck description~\cite{RMJ}\cite{RMP} 
was formulated to extract statistical properties of orbit ensembles and 
distribution functions over long time scales, assuming implicitly that the 
bulk potential is regular. To what extent, then, can Langevin 
realisations~\cite{VanK} of 
a Fokker-Planck equation yield reliable information about individual orbits 
over comparatively short time scales, particularly if the orbits are chaotic?

This is an issue of practical importance for systems like high intensity 
charged-particle beams. In older, low intensity beams, the contribution 
of the space charge to the total potential is comparatively minor compared 
to the confining magnetic field, but in high intensity beams the space charge
can become extremely important.
To what extent, then, is one justified in idealising the space charge by a
smooth density distribution that generates a smooth macroscopic potential?
Is it, {\em e.g.,} really legitimate to ignore discreteness effects entirely
in a bunch comprised of ${\sim}{\;}10^{9}-10^{10}$ protons with a transverse
emittance of a few microns?

The situation is especially suspect in that experience with `nearly 
collisionless' self-gravitating systems~\cite{HM} indicates that the
continuum limit must be subtle. One believes that, for 
$N\to\infty$, orbits in an $N$-body system will in fact converge towards 
characteristics in the corresponding smooth potential. However, $N$-body 
orbits evolved in a density distribution corresponding to an integrable 
potential are typically `very chaotic' with large positive Lyapunov exponents 
even though the integrable characteristics have vanishing Lyapunov exponent!

For flows in smooth potentials, it is straightforward to distinguish
macroscopically between phase mixing associated with regular versus chaotic
orbits~\cite{KN}. In particular, chaotic flows tend to mix exponentially
fast, whereas regular flows only mix as a more modest power law in time. 
Will this distinction persist in $N$-body systems? One might, {\em e.g.,}
worry that if $N$-body orbits corresponding to integrable characteristics
have large positive Lyapunov exponents, `regular' flows comprised of such
orbits would exhibit behaviour resembling chaotic phase mixing in smooth
potentials! Understanding the bulk properties of phase mixing in $N$-body
systems is important, {\em e.g.,} in light of recent grid code
simulations~\cite{Kis} indicating that chaotic phase mixing 
may be responsible for `anomalous relaxation'
observed in charged-particle beams, including the
University of Maryland `five beamlet' experiment~\cite{Rei}.

A complete resolution of these issues will require an analysis of `honest'
({\em i.e.,} direct summation~\cite{HE}) numerical integrations for systems 
comprised
of very large particle number $N$, which is not yet practical. However,
considerable insight may be derived by considering the simpler case of orbits
and orbit ensembles evolved in `frozen-$N$' systems, {\em i.e.,} 
time-independent $N$-body systems generated by randomly sampling a specified
smooth density distribution. In particular, by comparing orbits in such
frozen-$N$ systems with characteristics with the same initial condition evolved
in the corresponding smooth potential, one can quantify the
extent to which, as a function of $N$, $N$-body orbits and smooth potential
characteristics actually coincide. Such is the objective of this paper.


Section II focuses on the behaviour of orbits and orbit ensembles in
frozen-$N$ realisations of two simple potentials, one integrable and
the other almost completely chaotic. The principal conclusions here 
are (i) that there is a well-defined macroscopic convergence towards
the continuum limit, both for individual orbits and for orbit ensembles, 
and (ii) that discreteness effects can be extremely well-mimicked by
Gaussian white noise in the context of a Fokker-Planck or Langevin
description.

Section III considers the more realistic `thermal
equilibrium model'~\cite{BR}, well known from beam dynamics. For appropriate 
choices of parameter values, this model admits~\cite{BS} large measures of 
both regular and chaotic orbits, so that one encounters a new 
feature~\cite{BD}, namely transitions between regular and chaotic behaviour 
triggered by discreteness effects.
As for the simpler models considered in Section II, one observes clear
distinctions between regular and chaotic phase mixing, although discreteness
effects, again well modeled by a Fokker-Planck description, can be 
surprisingly important. Even when $N$ is large, individual orbits can 
exhibit frequent changes in behaviour corresponding macroscopically to 
transitions between regularity and chaos; and the scaling implict in a
Fokker-Planck description suggests that, for chaotic orbits, {\em even a total
particle number as large as $N{\;}{\sim}{\;}10^{9}$ may not be large enough 
to justify a continuum limit.}

Section IV focuses on Lyapunov exponents and the meaning of chaos in $N$-body
systems.
The principal conclusion here is that two distinct `types' of chaos can be 
present in the $N$-body problem, characterised by {\em two different sets of 
Lyapunov exponents associated with physics on different scales.}
Close encounters between particles trigger {\em microchaos}, a generic feature
of the $N$-body problem, which leads to large positive Lyapunov exponents
${\chi}_{N}$. If, however, the bulk smooth potential is chaotic, one also 
encounters {\em macrochaos}, which is again characterised by positive, albeit 
typically much smaller, Lyapunov exponents ${\chi}_{S}$. $N$-body realisations
of integrable systems remain chaotic, even for large $N$, in the sense that
${\chi}_{N}$ does not decrease towards zero for $N\to\infty$. 
Despite this, however, microchaos becomes progressively less important 
macroscopically in that the {\em range} of this chaos, {\em i.e.,} the scale 
on which the microchaos-driven exponential divergence of nearby orbits 
terminates, decreases with increasing $N$. 


Section V summarises the principal conclusions and speculates on potential
implications.
\section{$N$-Body Flows and $N$-Body Orbits in Regular and Chaotic Potentials}
\subsection{Models considered}
The computations described in this Section were performed for two models
which, albeit not representative of `real' equilibrium systems, are of 
significant pedagogical value in illustrating the nature of the continuum 
limit. In particular, since one model is integrable and the other almost
completely chaotic, it is simple to identify separately how discreteness 
effects impact regular versus chaotic orbits, an issue that becomes more
difficult in `realistic' systems which admit a complex coexistence of both 
regular and chaotic orbits.
\par\noindent
{\em Model 1.} A constant density triaxial ellipsoid, for which
\begin{equation}
{\rho}({\bf r})={3Q\over 4{\pi}abc} \times
\cases {  1 & if $m^{2}{\;}{\le}{\;}1$, \cr
               0     &  if $m^{2} > 1$, \cr}
\end{equation}
with
\begin{equation}
m^{2}= {x^{2}\over a^{2}} + 
{y^{2}\over b^{2}} + {z^{2}\over c^{2}} .
\end{equation}
For $m{\;}{\le}{\;}1$, this corresponds to a space-charge potential 
\begin{equation}
{\Phi}_{sc}({\bf r})={\Phi}_{0}-{1\over 2}\left(
{\omega}_{a}^{2}x^{2}+ {\omega}_{b}^{2}y^{2} + {\omega}_{c}^{2}z^{2} \right)
\end{equation}
with ${\Phi}_{0}$ a constant and frequencies ${\omega}_{a}$, ${\omega}_{b}$, 
and ${\omega}_{c}$ 
related to the axis values $a$, $b$, and $c$ in terms of incomplete 
elliptic integrals. The system was assumed confined by an external
potential ${\Phi}_{ext}=-2{\Phi}_{sc}$. It follows that, in the continuum
limit, each charge evolves in an integrable time-independent potential
of the form (modulo a constant)
\begin{equation}
{\Phi}_{reg}({\bf r})={1\over 2}\left(
{\omega}_{a}^{2}x^{2}+ {\omega}_{b}^{2}y^{2} + {\omega}_{c}^{2}z^{2} \right)
\end{equation}
Attention focused primarily on the parameter values 
$a=1.95$, $b=1.50$, and $c=1.05$, which, assuming units for which 
$Q{\;}{\equiv}{\;}1$, implies~\cite{Chan} that 
${\omega}_{a}{\;}{\approx}{\;}0.4663$,
${\omega}_{b}{\;}{\approx}{\;}0.5508$, and 
${\omega}_{c}{\;}{\approx}{\;}0.6753$.
It follows that the orbital time scale $t_{D}{\;}{\sim}{\;}2{\pi}/{\omega}
{\;}{\sim}{\;}10$.

Because this potential is integrable, one knows that, in the continuum
limit, only regular phase mixing is possible. However, the
situation is even more exceptional: because of the harmonic character of
the potential, {\em i.e.,} the fact that the force is linear, every charge 
will orbit with the same frequencies so that,
in the absence of discreteness effects, there can be no systematic phase 
mixing and no emittance growth in an initially localised clump. {\em All 
emittance growth associated with
this potential must be attributed to discreteness effects.}

\par\noindent
{\em Model 2.} Perturbing Model 1 by introducing a spherically symmetric, 
attractive spike of charge near the origin, which yields a modified potential
\begin{equation}
{\Phi}_{cha}({\bf r})={\Phi}_{reg}({\bf r})
-{q\over \sqrt{r^{2}+{\ell}^{2}}}
\end{equation}
with ${\ell}=10^{-3}$. Attention here focused on a central charge
$q=10^{-1.5}Q{\;}{\approx}{\;}0.03162$, which leads to a potential for 
which, for orbits restricted energetically to $m{\;}{\le}{\;}1$, {\em the 
phase space is almost completely chaotic}~\cite{KS02}. (The bulk properties
of the potential are insensitive to the precise value of ${\ell}$ for
${\ell}<10^{-2}$ or so; but most of the chaos is lost for much larger values of
${\ell}$.)

Frozen-$N$ charge density distributions of the form
\begin{equation}
{\rho}_{N}={1\over N}\sum_{i=1}^{N} {\delta}_{D}({\bf r}-{\bf r}_{i})
\end{equation}
were generated by randomly sampling the smooth density distributions
${\rho}$. These correspond to $N$-body potentials 
\begin{equation}
{\Phi}_{N}({\bf r})=-{1\over N}\sum_{i=1}^{N}
{1\over \sqrt{({\bf r}-{\bf r}_{i})^{2}+{e}^{2}}}
\end{equation}
which incorporate a tiny softening parameter $e$~\cite{sof}.
Unless otherwise stated, all Figures in this paper
were generated from integrations with ${e}=10^{-5}$.

Orbits were integrated in frozen-$N$ realisations with $N{\;}{\le}{\;}10^{6}$
using a variable time step scheme that conserved the energy of each charge
to at least one part in $10^{5}$. Estimates of the largest 
(finite time) Lyapunov exponent~\cite{GBP} were obtained in the usual way by 
simultaneously tracking the evolution of a small initial perturbation, 
periodically renormalised at fixed time intervals ${\Delta}t$. 

The efficacy of phase mixing was tested by generating localised clumps
of 1600 initial conditions sampling a
phase space region of size $|{\Delta}{\bf r}|{\;}{\sim}{\;}|{\Delta}{\bf v}|
{\;}{\sim}{\;}10^{-3}$ and evolving these into the future in both the smooth
potential and the corresponding $N$-body potential
(7). The resulting orbital data were analysed to compute emittances
\begin{equation}
{\epsilon}_{a}{\;}{\equiv}{\;}\sqrt{ {\langle}r_{a}^{2}{\rangle}
{\langle}v_{a}^{2}{\rangle}-{\langle}r_{a}v_{a}{\rangle}^{2}}
\qquad (a=x,y,z),
\end{equation}
as well as the total
\begin{equation}
{\epsilon}=({\epsilon}_{x}{\epsilon}_{y}{\epsilon}_{z})^{1/3}.
\end{equation}

The degree to which individual $N$-body orbits did, or did not, `look highly 
irregular' was quantified by computing  the {\em orbital complexity}~\cite{KEB}
of their 
Fourier spectra. This entailed determining for each orbit the quantities
$n_{x}$, $n_{y}$, and $n_{z}$, defined as the minimum number of frequencies
required to capture a fixed fraction $k$ of the power in each direction, and 
then assigning a total complexity
\begin{equation}
n(k)=n_{x}+n_{y}+n_{z}.
\end{equation}
In order to obtain a reasonably sharp Fourier spectrum, orbital data were
typically recorded at intervals ${\delta}t=0.05$, a time corresponding to 
less than 1\% of a typical orbital period.


\subsection{Regular and chaotic phase mixing}
In the continuum limit, initially localised clumps 
characterised
by the integrable potential (4) exhibit no systematic tendency to disperse. 
Because each orbit executes harmonic motions with the same three frequencies, 
the charges remain close together, returning to their original $x_{0}$,
$y_{0}$, and $z_{0}$, after periods ${\tau}_{x}$, ${\tau}_{y}$, an
${\tau}_{z}$. Discreteness effects break this exact periodicity
and trigger a systematic spread. This is illustrated in 
Figure 1, which exhibits the $x$ and $y$ coordinates of the same 1600 orbit 
ensemble with $E=1.0$ at five different times, allowing for frozen-$N$ 
backgrounds with
$N=10^{3.5}$ and $N=10^{5}$. For $N=10^{3.5}$, this dispersal
is comparatively rapid, the charges having spread to sample the entire
allowable configuration space within $t=128$, a time corresponding to 
only 10 orbital periods or so. For the larger system with $N=10^{5}$, the
dispersal is considerably slower, requiring a time $t{\;}{\sim}{\;}512$,
roughly four times larger, to achieve a comparable spread.

The situation is very different for the potential (5), for which, even in the
continuum limit, the particle phase space is almost completely chaotic. 
In this case, one observes exponentially fast chaotic phase mixing in the
smooth potential, and allowing for discreteness
effects only accelerates the process. This is 
evident from Figure 2, which exhibits the $x$ and $y$ coordinates for a
1600 orbit clump, again with $E=1.0$, evolved in frozen-$N$ realisations 
with $N=10^{3.5}$, $N=10^{4.5}$, and $N=10^{5.5}$, as well as (in the right
hand column) the smooth potential. Even for the smooth potential, a time
$t{\;}{\sim}{\;}128$ is sufficient for particles to sample most of the
energetically accessible phase space. 

The visual impression that the chaotic clump disperses far more 
rapidly can be quantified by computing the emittance
${\epsilon}$ as a function of time. The left hand column of Figure 3 
exhibits ${\epsilon}(t)$ for the same ensemble of initial conditions
used to generate Figure 1, now allowing for frozen-$N$ backgrounds with
$N=10^{3.0}$, $10^{3.5}$, $10^{4.0}$, $10^{4.5}$ and and $10^{5.0}$.
For the smallest value of $N$ it is not completely clear whether 
${\epsilon}$ grows exponentially or as a power law in time. However, 
for $N{\;}{\ge}{\;}10^{3.5}$, the growth is distinctly
subexponential. Indeed, the data for $N{\;}{\ge}{\;}10^{3.5}$ are well
fit by an emittance growth law
\begin{equation}
{\epsilon}{\;}{\propto}{\;}(t/t_{G})^{1/2}
\end{equation}
where 
\begin{equation}
t_{G}{\;}{\propto}{\;}Nt_{D}.
\end{equation}
Extrapolating to the limit $N\to\infty$ yields the expected 
result that there can be no systematic emittance growth.

The left hand column of Figure 4 exhibits analogous results for the initial
conditions used to generate Figure 2, now plotted on a logarithmic scale,
allowing for $N=10^{2.5}$, $10^{3.5}$, $10^{4.5}$, $10^{5.5}$, and, in 
the bottom panel, the smooth potential. It is evident that, for the
smooth potential, $\ln {\epsilon}$ exhibits a roughly linear growth
during the interval (say) $10<t<100$, corresponding to an exponential 
growth in emittance. This is hardly surprising. The fact that 
individual orbits in the clump are chaotic implies that they should
diverge exponentially so that, at least for small ${\epsilon}(0)$,
one would expect ${\epsilon}$ to grow exponentially at a rate comparable
to the value of a typical (finite time) Lyapunov exponent ${\chi}_{S}$
for the smooth potential. For this ensemble, the mean exponent for the
interval $0<t<256$ assumed the value ${\langle}{\chi}_{S}{\rangle}=0.056$,
which corresponds to the slope of the dashed line in panel (i). 

Allowing for discreteness effects clearly accelerates the rate of chaotic
phase mixing. For the two smaller values of $N$, $10^{2.5}$ and $10^{3.5}$,
the growth again appears exponential, albeit at a larger rate; but for
the systems with $N=10^{4.5}$ and $10^{5.5}$ the evolution is clearly more
complex. Indeed, a careful examination of the data for these two cases
suggests strongly that the evolution can be decomposed into two largely
distinct exponential phases, the former characterised by a growth rate
${\gg}{\;}{\chi}_{S}$ and the latter by a slower rate ${\sim}{\;}{\chi}_{S}$.
This is consistent with the analysis to be presented later in Section IV, 
which 
indicates that two sorts of chaos can act in $N$-body systems, large scale
{\em macrochaos} 
characterised by a Lyapunov exponent ${\chi}_{S}$ and shorter range {\em
microchaos} characterised by a Lyapunov exponent 
${\chi}_{N}{\;}{\gg}{\;}{\chi}_{S}$. (For very small $N$, the microchaos
also acts on macroscopic scales, thus overwhelming any observational effects
associated with the much weaker macrochaos: hence the (near-)absence of
the second exponential phase in panels (a) and (c)!) The data summarised 
in Figure 4 are consistent with a second exponential phase with
\begin{equation}
{\epsilon}{\;}{\propto}{\;}N^{-1/2}\exp({\chi}_{S}t),
\end{equation}
the form of which will be motivated in Section IId.

The results derived here for Model 2 which, in the continuum limit, 
corresponds to a chaotic potential, are likely generic for
bulk density distributions corresponding to a chaotic potential. However,
the results for Model 1 are special in that there is no
systematic emittance growth in the continuum limit. If, as one would expect
in a `real' system, the bulk potential exhibits at least some anharmonicities, 
regular phase mixing will trigger linear emittance growth even in the
continuum limit. In this case, allowing for discreteness effects will
again accelerate the growth of emittance, but the exact form of this
enhanced growth can be more complex, even though it will again be 
subexponential. Section III will exhibit additional examples of how 
discreteness can accelerate emittance growth for both regular and chaotic
clumps. 


\subsection{Individual Orbits and the Continuum Limit}
The preceding indicates that, as $N$ increases, orbit
ensembles evolved in frozen-$N$ backgrounds more closely resemble 
orbit ensembles with the same set of initial conditions evolved in the 
corresponding smooth potential. This does not, however, necessarily imply that 
individual orbits also converge towards characteristics in the smooth
potential. To what extent, then, is it true that, as $N$ increases, 
individual trajectories come to more closely resemble smooth potential 
characteristics? 

The most obvious -- and compelling -- check is visual: do frozen-$N$
orbits `look like' smooth potential characteristics when $N$ becomes 
sufficiently large? As a simple, and extreme, example, consider
a constant density spherical system without a central
spike where, in the continuum limit, ${\Phi}_{reg}$ reduces to an isotropic
harmonic oscillator potential; and select an initial condition which, in the 
smooth potential, corresponds to a circular orbit. Results derived from 
integrations of such an initial condition in different frozen-$N$ systems are
exhibited in Figure 5, which shows representative frozen-$N$ orbits generated 
for particle number varying between $N=10^{2.5}$ and
$N=10^{5.5}$, along with the smooth potential orbit. For the four smallest
values of $N$, there is no obvious hint that the orbit `should' be circular
or that there `should' be a net sense of circulation, although there {\em is} 
a crude visual sense that, as $N$ increases, the orbit becomes `less
tangled.' However, for $N=10^{4.5}$ one starts to discern a clear sense
of net circulation, for $N=10^{5.0}$ the orbit has clearly become
centrophobic (thus suggesting that angular momentum is at least approximately
conserved), and for $N=10^{5.5}$ the orbit arguably resembles a `distorted'
or precessing circular orbit. 

As is illustrated in Figure 6, the visual impression that the orbit is 
becoming more nearly circular can be corroborated by constructing the Fourier 
spectra of the orbital data. For the three smallest values of $N$, the
quantity $|x({\omega})|$ is obviously broad band, although there is a peak
at or near the circular frequency associated with the smooth potential
orbit. For $N=10^{4.0}$ and $10^{4.5}$ the peak becomes appreciably sharper,
and for $N=10^{5.0}$ and $10^{5.5}$ one sees only slight irregularities in
the spectra which translate into the visual appearance of precession.

The conclusion is obvious: as $N$ increases, frozen-$N$ orbits come to
more closely approximate the smooth orbit, both visually and in terms of
their power spectrum. Analogous results obtain for more generic initial
conditions evolved in this and other integrable potentials.

Convergence of orbits in terms of their Fourier spectra is important in
justifying straightforward applications of nonlinear dynamics to many-body
systems interacting via long range forces. Many physical phenomena in 
many-body systems, including accelerator modes~\cite{Chi}, modulational 
diffusion~\cite{Tenn}, and resonant relaxation~\cite{RT}, are attributed to 
resonant couplings between, {\em e.g.,} the natural frequencies of 
individual regular orbits and the frequency or frequencies associated with some
perturbation. However, such applications can only be justified if the `real'
$N$-body orbits have frequencies that adequately approximate the frequencies
associated with characteristics in the smooth potential.

The degree of irregularity exhibited by individual orbits can be quantified
by computing Fourier complexity, as defined in Eq.~(10). The results of
one such investigation are summarised by the curves with diamonds in Figure
7, which exhibit $n(0.95)$, the mean number of frequencies required to capture
95\% of the total power, for collections of 100 initial conditions evolved
in frozen-$N$ backgrounds with different $N$. (The triangles
will be discussed in the following subsection.) In each case, the initial
conditions were integrated for a time $T=128$, with orbital data 
recorded at fixed intervals ${\delta}t=0.05$. The data were then Fourier
analysed using an {\em FFT} solver 
to translate a set of $2j$ points into a set of $j$ Fourier amplitudes.
The upper panel was generated for orbits in the integrable Model 1, the 
lower panel for the strongly chaotic Model 2. In each panel,
the solid curve represents the mean complexity computed in the unperturbed
smooth potential. Error bars were computed by dividing the 100 initial
conditions in half and analysing each half separately.

It is evident that, in both cases, $n(0.95)$ is a decreasing function of $N$
which converges towards the continuum value for $N\to\infty$. For smaller
values of $N$, the regular and chaotic ensembles have comparable complexities,
although the regular ensemble is slightly less (${\sim}{\;}20\%$) complex.
However, for larger values of $N$ there are clear distinctions 
between the regular and chaotic ensembles, the chaotic ensembles for 
$N{\;}{\ge}{\;}10^{5}$
being nearly twice as complex as the regular ensembles. Indeed, for other
choice of regular models the complexity can be even lower: The fact that
the smooth potential complexity in panel (a) is as large as it is reflects
the fact that the regular orbits used to generate this Figure were relatively
complex `box' orbits, with the topology of Lissajous figures, which, even 
in the continuum limit, require two or three frequencies in each direction
to capture as much as 95\% of the power. If instead $n(0.95)$ is
computed for an ensemble of initial conditions corresponding to circular
orbits, the complexity converges towards a continuum limit with
$n(0.95)=3$.

\subsection{Modeling $N$-body orbits and flows by Gaussian white noise}
Conventional wisdom holds that discreteness effects can be idealised as 
friction and Gaussian (nearly) white noise in the context of a Fokker-Planck 
description~\cite{RMJ}. Taken literally, this suggests that individual 
$N$-body orbits can be well-mimicked by Langevin simulations. However, it is 
not completely clear to what extent this is really true. The original 
derivation of the Fokker-Planck equation (and most if not all of its tests) 
restricted attention to the statistical properties of orbit ensembles or
distribution functions over comparatively long time scales, assuming
implicitly that the bulk potential in which the particle evolves is 
nonchaotic. Can the friction/noise paradigm describe correctly
short time behaviour and/or the behaviour of individual orbits, especially
in a chaotic potential? 

Granted the validity of a Fokker-Planck description, a simple rule
connects $N$ to the strength of the friction and noise. Assuming that
the noise is characterised by a temperature per unit mass ${\Theta}$ 
comparable to the magnitude of the particle energy, the coefficient of 
dynamical friction ${\eta}$ defines an energy relaxation time
$t_{R}={\eta}^{-1}$. However, an evaluation of the Fokker-Planck coefficients
in a binary encounter approximation leads to the prediction that
$t_{R}{\;}{\propto}{\;}(N/ \ln {\Lambda})t_{D}$, with $\ln {\Lambda}$
the Coulomb logarithm. Given the assumption of a nonneutral plasma, the
treatment of ${\Lambda}$ must necessarily be somewhat heuristic~\cite{TON}.
However, there is a general agreement that ${\Lambda}$ should scale as some 
power of $N$, so that $t_{R}$, and hence 
the diffusion constant $D$, should satisfy 
\begin{equation}
t_{R}^{-1}{\;}{\propto}{\;}D{\;}{\propto}{\;}{\eta}{\;}{\propto}{\;}
{\ln N\over N}.
\end{equation}
The obvious question, then, is whether frozen-$N$ simulations with specified 
$N$ can be well-mimicked by Langevin simulations with
${\eta}{\;}{\propto}{\;}(\ln N /N)$.

Two practical issues arise in testing this prediction. The first is that, 
because of the limited range of $N$ that can be explored, it is impractical 
to test 
the subdominant $\ln N$ dependence: for $N<10^{3}$ or so, the very notion of 
a bulk potential fails; for $N>10^{6}$ computations become prohibitively 
expensive. One must instead restrict attention to testing the simpler scaling 
relation ${\eta}{\;}{\propto}{\;}N^{-1}$, {\em i.e.},
\begin{equation}
\ln {\eta}=p- \ln N,
\end{equation}
for some constant $p$.

The second point is more serious. The usual Langevin equation reads~\cite{VanK}
\begin{equation}
{d^{2}r_{a}\over dt^{2}}=-{\nabla}_{a}{\Phi}-{\eta}{dr_{a}\over dt}+F_{a},
\qquad (a=x,y,z),
\end{equation}
where ${\eta}dr_{a}/dt$ represents a dynamical friction. $F_{a}$ 
represents Gaussian white noise, which is characterised completely by its 
first two moments:
\begin{displaymath}
{\langle}F_{a}(t){\rangle}=0, \qquad (a,b=x,y,z)
\end{displaymath}
and
\begin{equation}
{\langle}F_{a}(t_{1})F_{b}(t_{2}){\rangle}=2{\eta}{\Theta}{\delta}_{ab}
{\delta}_{D}(t_{1}-t_{2}),
\end{equation}
with $D{\;}{\equiv}{\;}2{\eta}{\Theta}$ the diffusion constant 
entering into a Fokker-Planck description. By choosing ${\Theta}$ to equal the
initial energy one can ensure that the average energy of the orbits remains
unchanged. 

Such an equation is clearly unsatisfactory here. Energy is conserved 
absolutely for frozen-$N$ orbits, so that one must also impose energy 
conservation on any scheme which aims to mimic its effects. (For very small 
${\eta}$, energy remains almost conserved for very long times. However, 
comparatively small $N$ should correspond
to relatively large ${\eta}$, which implies large changes in energy and,
as such, significant changes in the phase space regions accessible to the
noisy orbit.) For this reason, the noisy integrations described here were
performed using a modified energy-conserving noise~\cite{SK}.

This entailed 
(1) eliminating the dynamical friction altogether, (2) again imparting random 
kicks as in Eq.~(17), but (3) renormalising the modified velocity at
each time step by an overall factor, {\em i.e.,} ${\bf v}(t+{\delta}t)\to
{\alpha}{\bf v}(t+{\delta}t)$, with ${\alpha}$ so chosen that 
$E(t+{\delta}t)=E(t)$. Modulo this complication, the noise was integrated
using a standard algorithm~\cite{GSH} based on a fourth order Runge-Kutta 
integration scheme with fixed time step ${\delta}t$. The integrations 
were performed for ${\delta}t=2\times 10^{-4}$, it having been confirmed that
the statistical effects of the noise were insensitive to the precise value
of ${\delta}t$ for ${\delta}t<10^{-3}$.

At the level of orbit ensembles, as probed by the emittance and other
bulk moments, the results of frozen-$N$ simulations are in fact extremely
well-mimicked by Langevin simulations, at least for comparatively large $N$.
The degree to which this is true can be inferred by contrasting the right
and left hand columns of Figures 3 and 4. As discussed already, the left
hand columns of Figures 3 and 4 exhibit, respectively, time-dependent 
emittances for Models 1 and 2, allowing for several different values of
$N$. The right hand panels exhibit data generated from Langevin integrations
of the same initial conditions, allowing for amplitudes ${\eta}$ satisfying
Eq.~(15) with $p=0.5$, so that, {\em e.g.,} $N=10^{5.5}$ corresponds to
${\eta}=10^{-5.0}$. For the smallest values of $N$ (and hence the largest 
values of ${\eta}$) -- corresponding to panels (a) and (b) in Figure 3 and
panels (a) - (d) in Figure 4, the agreement is not all that good. However,
for larger particle number -- $N{\;}{\ge}{\;}10^{3.5}$ for the regular system
and $N{\;}{\ge}{\;}10^{4.5}$ for the chaotic system --, the agreement is 
obviously quite good.

The bottom right hand panel in Fig.~4 was generated for orbits evolved with
a considerably smaller value of ${\eta}$, namely ${\eta}=10^{-7.5}$, this
corresponding to the largest noise amplitude that does {\em not} alter 
appreciably the emittance growth observed in the smooth potential. To the 
extent that the scaling of Eq.~(15) is in 
fact correct for $p{\;}{\approx}{\;}0.5$, the fact that larger values of
${\eta}$ have an appreciable effect on emittance growth implies that, even
over an interval as short as $t=128$, corresponding to ${\sim}{\;}10$ orbital
time scales $t_{D}$, one requires $N>10^{7}$ to justify a continuum limit! 
{\em Even though the collisional relaxation time scale 
$t_{R}{\;}{\propto}{\;}(N/ \ln N)t_{D}{\;}{\gg}{\;}t_{D}$,  
discreteness effects can be important in a system with 
$N{\;}{\sim}{\;}10^{6.5}$ on a time scale as short as ${\sim}{\;}10t_{D}$.}

As is evident from Figure 7,
this agreement also extends to the level of individual orbits.
As described already, the diamond curves in this Figure were derived from
frozen-$N$ integrations. The other curves, constituted of
triangles, were generated from exactly the same initial conditions, now
integrated, however, in the smooth potential while being subjected to
energy-conserving Gaussian noise with ${\Theta}=E$ and variable 
${\eta}$. To the extent that conventional Fokker-Planck theory is correct, 
one would 
anticipate a correspondence between $N$ and ${\eta}$ of the form given by
Eq.~(15). The noisy points in Figure 7 were in fact identified with the 
frozen-$N$ points assuming the validity of the scaling (15) with $p=0.6$. 
The obvious fact, then, is that, given this identification, the curves $n(N)$ 
and $n({\eta})$ rather nearly coincide. {\em Even at the level of the
complexity of individual orbits, frozen-$N$ orbits can be well-mimicked by
noisy orbits with} $\ln {\eta}+\ln N=$ const.


Granted that discreteness effects can be mimicked by Gaussian white 
noise, the scaling relations (11) and (13) are easily understood.
At least for a harmonic potential, it is 
straightforward to derive analytic solutions to the Langevin equation (16) for 
moments like ${\langle}x^{2}{\rangle}$ or 
${\langle}xv_{x}{\rangle}$~\cite{RMP}.
Alternatively, it is easily seen that the Fokker-Planck equation associated 
with Eq.~(16) implies that the clump emittance satisfies
\begin{equation}
{d{\epsilon}_{x}^{2}\over dt}=2{\eta}{\Theta}+2{\eta}{\Bigl(}
{\langle}x^{2}{\rangle}{\langle}v^{2}{\rangle}-{\langle}xv{\rangle}^{2}
{\Bigr)}.
\end{equation}
Assuming, however, that the initial emittance is extremely small, at early
times one can approximate that 
${\langle}x^{2}{\rangle}{\langle}v^{2}{\rangle}{\;}{\approx}{\;}
{\langle}xv{\rangle}^{2}$; and, to the extent that the growth time 
is long compared with the characteristic crossing time, one can average over
oscillations to set
${\langle}x^{2}{\rangle}=E/{\omega}^{2}$, with $E$ the initial energy.
It then follows that, for early times, 
\begin{equation}
{\epsilon}_{x}{\;}{\approx}{\;}{\Bigl(}
{2E{\Theta}{\eta}t\over {\omega}^{2}}{\Bigr)}^{1/2}.
\end{equation}
Combining Eq.~(19) and the analogous formulae for ${\epsilon}_{y}$ and 
${\epsilon}_{z}$ with the scaling relation
${\eta}{\;}{\propto}{\;}1/N$ leads immediately to Eq.~(11).
The same diffusive $t^{1/2}$ behaviour also arises for ${\delta}r_{rms}$ and
${\delta}v_{rms}$. 

A somewhat more heuristic argument can account for the scaling (13) 
associated with a chaotic potential. If the initial emittance 
${\epsilon}(0)=0$, it is clear that, in the absence of discreteness effects, 
${\epsilon}(t)$ would continue to vanish identically: two smooth integrations
of the same initial condition will not yield divergent orbits, even if the
orbits are chaotic. However, discreteness effects act to `kick' two
nearly coincident orbits apart, at which point they will tend to diverge at a
rate set by the Lyapunov exponent ${\chi}_{S}$ associated with the bulk
potential. Assuming, however, that the `kicks' are random, their effects will 
scale as ${\eta}^{1/2}$ rather than ${\eta}$; but combining this with
Eq.~(15) implies that~\cite{HKM} 
\begin{equation}
{\delta}r_{rms}{\;}{\propto}{\;}{\delta}v_{rms}{\;}{\propto}{\;}
N^{-1/2}\exp({\chi}_{S}t).
\end{equation}
Eq.~(13) follows since ${\epsilon}$ scales as ${\delta}r_{rms}$ and
${\delta}v_{rms}$.
\section{The Thermal Equilibrium Model}
\subsection{Defining the model}
Consider now a more realistic example, the thermal equilibrium model~\cite{BR},
which, in the continuum limit, admits large measures of both regular and
chaotic orbits~\cite{BS}. This model allows for a collection of $N$ identical
charged particles, interacting electrostatically, that is constrained by 
linear restoring forces to manifest triaxial symmetry, the focusing
forces in different orthogonal directions being characterised in general by
unequal frequencies. Individual particles thus have energy
\begin{equation}
E={1\over 2}mv^{2}+{1\over 2}m({\bf\omega}{\cdot}{\bf x})^{2}+
q{\phi}({\bf x}),
\end{equation}
where ${\bf x}$ and ${\bf v}$ denote particle position and velocity,
$m$ and $q$ denote the mass and charge, 
${\bf\omega}=({\omega}_{x},{\omega}_{y},{\omega}_{z})$
represents the three frequencies associated with the focusing force,
and ${\phi}({\bf x})$ is the collective space-charge potential. 

The additional assumption is that the particles can be characterised by a 
one-particle distribution function of the Maxwell-Boltzmann form,
$f{\;}{\propto}{\;}\exp(-E/k_{B}T)$, with $k_{B}T$ the temperature. 
This implies a bulk number density satisfying
\begin{equation}
n({\bf x})=n(0)\exp \left[
{-{1\over 2}m({\bf\omega}{\cdot}{\bf x})^{2}-q{\phi}({\bf x}) \over
k_{B}T} \right]
\end{equation}
where ${\phi}({\bf x})$ is defined implicitly as a function of $n$ via 
the relations (in {\rm mks} units)
\begin{equation}
{\nabla}^{2}{\phi}({\bf x})=-{q\over {\varepsilon}_{0}}n({\bf x}),
\qquad {\phi}(0)={\nabla}{\phi}(0)=0.
\end{equation}

Following, {\em e.g.,} \cite{BS}, the problem can be cast into dimensionless
form by expressing length and frequency in units of the Debye length and plasma
frequency, {\em i.e.},
\begin{equation}
{\lambda}_{D}^{2}={{\varepsilon}_{0}k_{B}T\over n(0)q^{2}} , \qquad
{\omega}_{p}^{2}={n(0)q^{2}\over {\varepsilon}_{0}m},
\end{equation}
and by introducing a dimensionless potential
\begin{equation}
{\Phi}({\bf x})={q{\phi}({\bf x})\over k_{B}T}.
\end{equation}
With appropriate rescaling, the net result is a density distribution of
the form
\begin{equation}
n({\bf x})=\exp \left[
-{1\over 2}{\Omega}^{2}R^{2}({\bf x})-{\Phi}({\bf x}) \right],
\end{equation}
where
\begin{equation}
{\nabla}^{2}{\Phi}({\bf x})=-n({\bf x}),
\qquad {\Phi}(0)={\nabla}{\Phi}(0)=0.
\end{equation}
Here ${\Omega}^{2}=({\omega}_{y}/{\omega}_{p})^{2}$
and $R^{2}=(x/a)^{2}+y^{2}+(z/c)^{2}$, in terms of scale lengths $a$ and $c$
satisfying $a={\omega}_{y}/{\omega}_{x}$ and $c={\omega}_{y}/{\omega}_{z}$ .
The minimum permissible focusing strength corresponds to 
\begin{equation}
{\Omega}={\Omega}_{u}=1/\sqrt{(1/a^{2})+1+(1/c^{2})}.
\end{equation}
The experiments described here were performed assuming parameter values
$a^{2}=0.5$, $c^{2}=1.5$, and ${\Omega}=1.0001/\sqrt{3}$, for which a typical 
orbital time scale $t_{D}{\;}{\sim}{\;}20$.


These parameters represent a beam that is moderately, but not strongly,
dependent on space charge. Consider, for example, a proton bunch with 
1 $\mu$m root mean squared normalised emittance spanning 3 cm full `radius'.
If the bunch is described by the thermal equilibrium model,
the Debye length is ${\sim}{\;}2$ mm and the bunch population is the 
${\sim}{\;}3\;\times\;10^9$ protons, this 
corresponding to a bunch charge ${\sim}{\;}0.5$ nC.

In general it does not appear possible to solve eqs.~(26) and (27) 
analytically. This makes both the generation of $N$-body realisations of the 
density and the computation of orbits in the smooth potential much more
difficult. However, these difficulties can be, and were, resolved using 
numerical techniques described in~\cite{BS}. 
In principle, the accelerations for the N-body thermal equilibrium model 
should be given by 
\begin{equation}
\bar{a}=-\nabla{\Big[}\frac{1}{2}\Omega^2R^2+\frac{1}{4\pi}\frac{N}{N_m}
\sum_{i=1}^{N}{1\over \sqrt{
|{\bf r}-{\bf r}_{i}|^{2}+e^{2}}}{\Big]}
\end{equation}
with $N$ the number of frozen particles and $N_m$ satisfying 
\begin{equation}
N_m=\int_{-\infty}^{\infty}\,d^{3}r
\exp\left[ {1\over 2}{\Omega}^{2}R^{2} - {\Phi}({\bf r})\right].
\end{equation}
In practice, however, one cannot perform this integral, even numerically,
since ${\Phi}$ was only evaluated on a finite grid. For this reason, the
integral was first solved with limits coinciding with the grid boundaries,
but then renormalised by a small constant factor so that
plots of the potential and density in the smooth and frozen-$N$ configurations
overlapped perfectly.

\subsection{Regular and chaotic phase mixing}
\par\noindent
The top four left hand panels of Fig.~(8) exhibit emittance growth for an
initially localised orbit ensemble evolved in frozen-$N$ realisations of
the thermal equilibrium model, selected 
with energy sufficiently small that the constant energy hypersurface in the
smooth potential is completely regular. Since the size of the accessible 
phase space is roughly ten times larger than was the case for the 
oscillator models, the initial conditions sampled a region ten times larger,
$|{\Delta}r|{\;}{\sim}{\;}|{\Delta}v|{\;}{\sim}{\;}10^{-2}.$ 
 It is evident that, as for the
integrable oscillator model considered in Section II, the emittance growth
is power law rather than exponential; and, at least for $N=10^{5.5}$ and 
$N=10^{6.0}$, it is well fitted by a growth law 
${\epsilon}{\;}{\propto}{\;}t^{1/2}$. 

The bottom left panel exhibits emittance growth for the same orbit ensemble
evolved in the smooth potential. Here the evolution is clearly linear, rather
than square root, the expected behaviour for smooth orbits in generic 
integrable potentials where nearby initial conditions correspond to slightly 
different orbital frequencies. (Recall that, for the oscillator Model 1, 
there is zero emittance growth in the continuum limit.) That 
${\epsilon}$ grows much faster for the frozen-$N$ model with $N=10^{6.0}$
than for orbits in the smooth potential indicates clearly that, for the
thermal equilibrium model, $N=10^{6.0}$ is not a sufficiently large particle
number to justify a continuum approximation, even over an interval as short
as $t=512$. 

The top four right hand panels demonstrate that discreteness effects can 
again be well-mimicked by energy-conserving Gaussian white noise with 
${\Theta}=E$ and 
$N$ and ${\eta}$ related as in Eq.~(15) although, in this case, the best fit 
value $p{\;}{\approx}{\;}1.5$, rather than $p{\;}{\approx}{\;}0.5$. 
The bottom right panel exhibits the emittance growth for an ensemble evolved
with ${\eta}=10^{-6.5}$, the largest noise amplitude that does not 
significantly alter emittance growth in the smooth potential. Presuming that 
the scaling (15) can again be extended to larger $N$ and smaller ${\eta}$, 
one infers that, for the purpose of predicting emittance growth in this 
regular ensemble, the smallest value of $N$ for which the 
continuum limit can be justified is $N{\;}{\sim}{\;}10^{8}$. 

Figure 9 exhibits analogous data for a higher energy ensemble which, in 
the continuum limit, corresponds completely to chaotic orbits. It is evident
that, as for the chaotic model in Section II, the evolution is exponential
overall, rather than power law; and that discreteness effects again have an
important effect. Also evident from a comparison of left and right hand
panels is that, as for regular orbits, discreteness effects can be 
well-mimicked by noise with $N{\;}{\propto}{\;}1/{\eta}$ and 
$p{\;}{\approx}{\;}1.5$. Most striking, however, is the fact that, in this
case, even much weaker noise can accelerate emittance growth appreciably. 
For this chaotic ensemble, 
discreteness effects must correspond to a
noise amplitude satisfying ${\eta}<10^{-8.0}$ or so before a continuum 
limit can be justified. {\em Chaotic orbits are far more susceptible to low 
amplitude noise than are regular orbits.} Presuming again that the scaling 
relation (15) holds, this implies that {\em for the case of chaotic orbit 
ensembles, the continuum limit cannot be justified for $N<10^{9.5}$.}

This, coincidentally, is roughly the number of particles in the equilibrium 
proton bunch described in Sec. III A. Accordingly, in studying the dynamics 
of beams with moderate space charge, one may not be able to assume the 
validity of the continuum limit with complete confidence, even for a system 
in equilibrium. The situation may be even more problematic for a beam that
is significantly out of equilibrium, since the resulting time-dependent
potential would be expected to generate a larger population of chaotic 
orbits~\cite{Spr}.

\subsection{Transitions between regular and chaotic behaviour}
At very low energies, where the total potential is nearly harmonic,
all smooth potential orbits are regular, so that discreteness effects can only
act to deflect frozen-$N$ orbits from one regular trajectory to another.
However, at higher energies the smooth potential admits a complex coexistence
of regular and chaotic orbits. This implies the possibility that discreteness 
effects can deflect frozen-$N$ orbits from regular to chaotic characteristics
and vice versa~\cite{BD}. Of obvious interest then is how fast, as a function
of $N$, such transitions occur. 

For example, an accelerator designer relying on the Vlasov equation and
an analysis based on a smooth, macroscopic potential would neglect these 
microscopic transitions and their corresponding impact on chaotic mixing.  
Thus the physics of collective relaxation and global emittance growth would 
be improperly modeled and the results, at least in principle, would be
suspect.  The consequences of such an omission depend on the problem at
hand, but one might expect them to be especially severe for nonequilibrium 
beams where chaotic dynamics is likely to be more prevalent~\cite{Spr}.

If one selects a localised ensemble of initial conditions corresponding to 
regular orbits in the smooth potential and integrates these initial conditions
into the future, discreteness effects will, if sufficiently strong, eventually 
trigger transitions from regularity to chaos. That such transitions actually 
occur can be determined by a visual inspection of individual frozen-$N$ orbits 
which can be observed to become abruptly `more irregular' in appearance. If, 
moreover, large numbers of transitions occur over very short times, this can
make the emittance associated with an initially localised ensemble, which 
ought to grow as a power law in time, exhibit instead a more rapid, roughly 
exponential, increase. 

However, an accurate determination of the relative fraction of the orbits 
which are still regular requires a more careful analysis.
This was done by recording the phase space coordinates of individual frozen-$N$
orbits at various times $t>0$, and evolving these into the future in the
smooth potential to determine whether the resulting smooth potential
characteristics were still regular or whether instead they had become chaotic. 
The most straightforward fashion in which to determine whether the orbits are 
chaotic would have been to compute an estimate of the largest (finite time) 
Lyapunov exponent. Given, however, that the potential cannot be expressed in
terms of elementary functions, this would have proven extremely expensive 
computationally. For that reason, distinctions between regularity
and chaos were based instead on the computed {\em complexities} of the
characteristics. As discussed elsewhere ({\em e.g.,} \cite{BS}, \cite{KEB},
and references cited therein), such a criterion typically coincides almost
exactly with more conventional criteria based on Lyapunov exponents.

Presuming that the system is ergodic and that discreteness effects are 
sufficiently strong that they can in principle convert any orbit from regular
to chaotic, and vice versa, it would seem clear what such an analysis ought
to reveal. (1) At sufficiently late times, independent of $N$ the relative 
fraction of chaotic orbits generated from any initial ensemble should (to 
within statistical uncertainties) coincide with the relative measure of chaotic
orbits on the constant energy hypersurface, {\em i.e.,} to the relative volume 
of the chaotic portions of the constant energy hypersurface. (2) Assuming,
however, that discreteness effects are more important for smaller $N$, the
time required to converge towards this asymptotic value should be an increasing
function of $N$. As $N$ increases, transitions should become more rare.

As illustrated in Fig.~10, which summarises computations with particle
number between $10^{4.5}$ and $10^{6.0}$, this expectation was in fact 
confirmed. For frozen-$N$ systems with number as small as
$N=10^{4.5}$, nearly 40\% of the orbits in an initially regular ensemble 
had become chaotic within a time $t=64$, a time corresponding to only
${\sim}{\;}3t_{D}$, and the relative measure $f$ of 
chaotic orbits appears to have asymptoted towards a time-independent value by 
$t=128$. The relative measure of chaotic orbits for computations with 
$N=10^{5.0}$ grew more slowly in time; but, by $t=512$, the relative measure
had again approached a value $f{\;}{\sim}{\;}0.4$. For larger values of 
$N$, $f$ remains a monotonically increasing function of time, but transitions 
are sufficiently rare that one does not approach an equilibrium population
within a time as short as $t=512{\;}{\sim}{\;}25t_{D}$.

\section{Lyapunov Exponents for Microchaos and Macrochaos}
\subsection{Ordinary Lyapunov exponents}


As $N$ increases, frozen-$N$ orbits come to more closely
resemble smooth potential characteristics generated from the same initial
condition, both visually and in terms of their Fourier spectra.  One might 
therefore expect that, at least for a regular, {\em i.e.,} nonchaotic,
smooth potential, the value of the
largest Lyapunov exponent ${\chi}_{N}$ should decrease with increasing $N$ and 
converge towards zero for $N\to\infty$. Such, however, is not the case. Rather,
as is also true for gravitationally interacting systems of particles~\cite{SK},
for both regular and chaotic orbits 
the value of ${\chi}_{N}$ is comparatively insensitive to $N$; and there are 
even indications that ${\chi}_{N}$ might {\em increase} with increasing $N$.


Figure 11 exhibits the value of the largest Lyapunov exponent ${\chi}_{N}$
as a function of $N$ for a single initial condition evolved in frozen-$N$
integrations with softening parameters varying between 
${e}=10^{-5}$ and ${e}=10^{-1}$. 
Figure 12 exhibits the same data, now plotting ${\chi}_{N}$ as a function
of ${e}$ for different values of $N$. In both Figures, each point was 
generated by
integrating the same initial condition used to generate Figures 5 and 6 for 
a total time $t=256$ in 10 different frozen-$N$ realisations of Model 1.
In the continuum limit this initial condition corresponds to an integrable
circular orbit with vanishing Lyapunov exponents; and, as was evident from 
Figure 5, the frozen-$N$ orbits for larger $N$ look much more regular 
in appearance than do the orbits with smaller $N$. Despite this, however, 
at least for the smallest values of ${e}$, 
${\chi}_{N}$ does not decrease with increasing $N$. As probed by ${\chi}_{N}$, 
orbits with $N=10^{2.5}$ and $10^{5}$ are comparably chaotic!

However, for larger values of 
${e}$, ${\chi}_{N}$ {\em does} decrease with increasing $N$. That
this should be the case is easily understood. Because the bulk potential 
is integrable, the chaos must at some level be associated with close 
encounters between nearby charges; but introducing a 
softening parameter {\em de facto} `turns off' encounters on scales shorter
than ${e}$. If the charge density
is sufficiently large that encounters with separation $<{e}$ become
common, the simulation will have artificially, and erroneously, reduced
this source of chaos, yielding a smaller ${\chi}_{N}$. 

That the value of ${\chi}_{N}$ for nearly unsoftened frozen-$N$ integrations
is insensitive to whether the smooth potential is regular or chaotic is
illustrated in Figure 13, which exhibits ${\chi}_{N}$ as a function of
$N$ for integrations with ${e}=10^{-5}$ for the same initial condition
integrated in both Models 1 and 2. That Figure also emphasises another
important point, namely that ${\chi}_{N}$ is typically larger than any 
Lyapunov exponent ${\chi}_{S}$ associated 
with motion in the smooth potential by an order of magnitude or more. For 
this particular initial condition, ${\chi}_{N}{\;}{\approx}{\;}0.82$ whereas 
${\chi}_{S}{\;}{\approx}{\;}0.056$. 

That ${\chi}_{N}$ should be insensitive to the choice of $N$, at least for
unsoftened simulations, might seem rather curious. However, it is not hard to 
understand why this might be so for a $1/r^{2}$ force.
Given that the microchaos disappears
completely in the continuum limit, it would seem clear that it must be 
associated with a sequence of `random' interactions between a `test' charge
and a collection of `field' charges. However, this suggests that the Lyapunov
time $t_{*}{\;}{\equiv}{\;}{\chi}_{N}^{-1}$ associated with the growth of a 
small initial perturbation can be estimated by considering tidal effects 
associated with a pair of charges separated by a distance $r$ comparable to 
(some fixed fraction of) the typical interparticle separation $r_{sep}$.
This tidal acceleration scales as 
\begin{equation}
{d^{2}{\delta}{\bf r}\over dt^{2}}=({\delta}{\bf r}{\cdot}{\nabla}){\bf a}
{\;}{\propto}{\;}{q\over r_{sep}^{3}}{\delta}{\bf r}{\;}{\equiv}{\;}
{{\delta}r\over t_{*}^{2}},
\end{equation}
with $q$ the magnitude of an individual charge. Given, however, that 
$r_{sep}{\;}{\sim}{\;}n^{-1/3}{\;}{\sim}{\;}N^{-1/3}R_{sys}$,
with $R_{sys}$ the size of the system and $n$ a characteristic number density, 
and that $q=Q/N$, with $Q$ the total charge, it follows that the time scale 
$t_{*}$, and hence ${\chi}_{N}$, should be independent of $N$. As $N$ 
increases, the sizes of the individual charges and the cube of the typical 
interparticle separation both decrease as $N^{-1}$, so that the ratio is 
independent of particle number.

A more careful argument~\cite{Pogo} actually allows one to prove analytically
that, in the absence of softening, ${\chi}_{N}$ cannot decrease towards
zero with increasing $N$. For simple geometries, an analytic expression for 
the average value of the stability matrix entering into the definition of 
${\chi}_{N}$ can be formulated in terms of a $3N$-dimensional 
integral. This integral cannot be evaluated analytically, but one {\em can}
derive rigorous bounds which ensure that the largest eigenvalue 
remains positive even for $N\to\infty$; and, even more strikingly, 
Monte Carlo evaluations of the integrals suggests that ${\chi}_{N}$ should be a
slowly {\em increasing} function of $N$. In other words, viewed in terms
of ${\chi}_{N}$, orbits become more chaotic as $N$ increases, even though,
for the case of a regular potential, they become more regular in appearance!
\cite{Foot}

{\em The obvious inference is that $N$-body Lyapunov exponents ${\chi}_{N}$ do
not provide a useful characterisation of the degree of chaos associated with
an orbit, at least when that orbit is viewed macroscopically.}

\subsection{Microchaos and macrochaos in the $N$-body problem}
That frozen-$N$ orbits have large positive 
Lyapunov exponents ${\chi}_{N}$, even for the case of an integrable potential,
but that distinctions between regular and chaotic potentials are clearly
manifested in the phase mixing of initially localised clumps  would 
suggest that there are two different, and comparatively distinct, potential
sources of chaos in the $N$-body problem.

On the one hand, one would expect {\em microchaos}, triggered by close 
encounters between individual charges, which is manifested only on very
short scales, comparable to, or perhaps somewhat larger than, the typical 
interparticle spacing. This source of chaos should be generic to the $N$-body
problem, irrespective of the form of the bulk potential, generating 
randomness qualitatively similar to what arises in pinball.
On the other hand, there is also the possibility of larger scale
{\em macrochaos}, which would be expected if and only if the bulk potential 
admits global stochasticity.

If these expectations are in fact correct, two nearby
initial conditions evolved in a frozen-$N$ realisation of any potential
should diverge exponentially at a rate ${\sim}{\;}{\chi}_{N}$ until their
separation becomes somewhat larger than a typical interparticle spacing, 
at which point the microchaos would `turn off.' If the bulk potential is
regular, no other source of chaos could act and the two orbits would continue
to diverge as a more modest power law. If, however, the bulk potential is
chaotic, {\em macrochaos} would still act, resulting in a continued 
exponential divergence, albeit at a rate ${\sim}{\;}{\chi}_{S}$ 
typically much smaller than ${\chi}_{N}$. 
In this case, exponential
divergence should be replaced by a power law divergence only once the 
separation has become macroscopic.

This three-stage evolution for chaotic orbits is clearly illustrated in
Figure 14, which exhibits data for frozen-$N$ simulations with $N=10^{5.0}$,
$10^{5.5}$, and $10^{6.0}$. The three curves in that figure were each
generated by selecting $50$ fiducial initial conditions in a phase space 
region of size $5\times 10^{-4}$ and $50$ perturbed initial 
conditions 
that were subjected to a displacement ${\delta}x=10^{-5}$, evolving each
initial condition into the future and computing as a function of time
the phase space separation~\cite{LL}
\begin{equation}
{\delta}Z=\sqrt{|{\delta}{\bf r}|^{2}+|{\delta}{\bf v}|^{2}}.
\end{equation}
That ${\delta}Z$ experiences two distinct stages of exponential evolution,
at very different rates is especially evident in the curve with
$N=10^{6.0}$. The solid lines accompanying that curve have slopes
$0.82$ and $0.056$, corresponding, respectively, to the mean values of 
the $N$-body ${\chi}_{N}$ and the smooth potential ${\chi}_{S}$ for those
initial conditions. 

As $N$ increases, the initial exponential phase terminates for smaller values 
of ${\delta}Z$. The microchaos responsible for this first phase
will `turn off' when $|{\delta}{\bf r}|$ becomes large compared with a
typical interparticle spacing, but that interparticle spacing 
scales as $N^{-1/3}$. 

Figure 15 shows the analogue of Figure 14, now generated for regular initial
conditions in the thermal equilibrium model. Once again there is an initial 
exponential growth 
at a rate comparable to ${\chi}_{N}$, but in this case there is no evidence 
of a second exponential phase. Rather, the initial exponential phase is
followed immediately by an interval of power law divergence. 

One other point, not obvious from these Figures, is that the scaling
with $N$ observed for the final power law phase is different for the regular
and chaotic systems. For regular orbits, the phase space separation
${\delta}Z$, like  $|{\delta}{\bf r}|$ and $|{\delta}{\bf v}|$, satisfies
a linear growth law
\begin{equation}
{\delta}Z(t){\;}{\propto}{\;}(t/t_{G}),
\end{equation}
with 
\begin{equation}
t_{G}{\;}{\propto}{\;}N^{-1/2}t_{D}.
\end{equation}
For chaotic orbits, one finds that ${\delta}Z$ again grows linearly
in time, but that the growth time
\begin{equation}
t_{G}{\;}{\propto}{\;}(1/\ln N)t_{D}.
\end{equation}


\subsection{Alternative interpretations of Lyapunov exponents}
The standard definition of Lyapunov exponents implies that they probe the
average rate of divergence for two nearby chaotic orbits in a single system.
However, for the $N$-body problem, Lyapunov exponents also 
quantify two other effects which, as a practical matter, are of equal 
importance:
\par\noindent
1. {\em Lyapunov exponents probe the rate at which orbits generated from the
same initial condition but evolved in two different frozen-$N$ systems diverge 
from one another.}
\par\noindent
2. {\em Lyapunov exponents probe the rate of divergence associated with 
orbits evolved from the same initial condition in both a 
frozen-$N$ system and in the smooth potential.}

This means that Lyapunov exponents also provide information about
the degree to which characteristics generated in the smooth potential can
be interpreted as providing a pointwise approximation to real frozen-$N$
orbits with the same initial condition, as well as the degree to which orbits 
in two different frozen-$N$ systems remain close to one another in a pointwise 
sense. 

This is important
at a practical level. In a real experiment one may perhaps be able to ensure 
that a given $N$-body system constitutes (nearly) a fair sampling of some 
specified density distribution, but the details of the actual $N$-body
distribution are likely inaccessible. Of obvious interest, therefore, are
the questions: to what extent will orbits in two different $N$-body systems
coincide? and to what extent do such orbits coincide with characteristics
in the bulk potential associated with the smooth density distribution?

Figure 16 exhibits the analogue of Figure 14, now generated by comparing
the same 50 initial conditions evolved in two different frozen-$N$ 
systems. Figure 17 compares orbits in a frozen-$N$ system with orbits in
the smooth potential. In each case, the duration of the initial interval
of especially fast exponential divergence is significantly
reduced, but the second interval with divergence at a rate 
${\sim}{\;}{\chi}_{S}$ is still conspicuous.

It is not hard to understand why the smooth potential 
${\chi}_{S}$ provides information about orbits in different frozen-$N$ 
simulations and/or their relation to orbits in the smooth potential. 
As noted already, discreteness effects can be extremely well-mimicked by
noise, at least mesoscopically. However, after the rapid decay of any initial 
transients, multiple noisy realisations of the same initial condition 
corresponding to a chaotic orbit typically diverge exponentially in such 
a fashion that~\cite{HKM}
\begin{equation}
{\delta}Z{\;}{\propto}{\;}({\Theta}{\eta})^{1/2}\exp({\chi}_{S}t)
{\;}{\propto}{\;}N^{-1/2}\exp({\chi}_{S}t),
\end{equation}
where the second proportionality follows from the observed scaling
${\eta}{\;}{\propto}{\;}1/N$.

\section{Conclusions}
Viewed macroscopically, there is a precise sense in which, as $N$ increases, 
trajectories in frozen-$N$ systems converge towards characteristics in the 
corresponding smooth potential. For very small particle number, $N<10^{4}$ 
or so, the notion of an average bulk potential fails and orbits in frozen-$N$
systems are very different from smooth potential characteristics. In
particular, the usual distinctions between regularity and chaos that exist
in a smooth potential seem completely lost.~\cite{smo} However, for larger
$N$ one begins to observe clear distinctions between orbits evolved from
initial conditions which, in the continuum limit, correspond to regular versus
chaotic orbits. 

In particular, although discreteness effects cannot be
neglected, phase mixing of initially localised orbit ensembles in a 
frozen-$N$ environment allows for clear distinctions between `regular' and 
`chaotic' clumps. Just as for clumps evolved in a smooth potential, 
emittance growth for a regular frozen-$N$ clump proceeds as a power law
in time, whereas it is roughly exponential for a chaotic clump. However, 
in both cases the growth is more rapid than in the smooth potential.
{\em Discreteness effects accelerate emittance growth for both regular and
chaotic clumps.}

In terms of both the statistics of orbit ensembles and the complexity of
individual orbits, discreteness effects can be extremely well-mimicked by
Gaussian white noise in the context of a Fokker-Planck, or Langevin, 
description, with a coefficient of dynamical friction ${\eta}$ and a diffusion 
constant $D$ consistent with the predicted scaling
$D{\;}{\propto}{\;}{\eta}{\;}{\propto}{\;}(\ln {\Lambda})/N$,
with $\ln {\Lambda}$ the Coulomb logarithm. A Fokker-Planck/Langevin 
description appears justified even when considering the short time behaviour 
of individual orbits. This suggests strongly that {\em Langevin simulations can
be used to assess the importance of discreteness effects in systems where
$N$ is too large to allow honest direct summation integrations.} 

To the extent that such an extrapolation is justified, one concludes that
discreteness effects can remain important even for very large $N$, especially
for the case of chaotic orbits. Consider, {\em e.g.,} the role of discreteness
effects in accelerating  emittance growth for an initially localised 
clump. For the case of the thermal equilibrium model, a relatively benign 
system without particularly large density contrasts and without internal 
substructures, one needs $N=10^{8}$ or more to justify the continuum 
approximation in tracking the evolution of a regular clump confined initially
to a region ${\sim}{\;}10^{-3}$ the size of the accessible phase space. For the
case of a chaotic clump 
of comparable size, one needs at least $N=10^{9.5}$.
This has obvious implications for beams in that it affects macroscopic 
mixing and associated changes in the overall phase-space volume.
Discreteness effects are also important because they can trigger transitions 
between regular and chaotic behaviour, a potentially serious problem for
charged particle beams. One might, {\em e.g.,}, try to initialise a bunch
in such a fashion that, although the bulk potential admits chaotic orbits,
only regular regions are populated, thus aiming to facilitate emittance 
compensation. The problem, however, is that discreteness effects could
transform significant numbers of orbits from regular to chaotic, thus making
compensation far more difficult. 

The time scale associated with transitions between regularity and chaos
increases with increasing $N$, such transitions being impossible in the
continuum limit; but for any finite $N$ there is presumably a maximum time
over which it is safe to ignore these transitions. The critical point, then,
as is evident from Fig.~10, is that that time can be much shorter than the 
collisional relaxation time $t_{R}$. To the extent that discreteness effects
in the thermal equilibrium model can be mimicked by Gaussian white noise,
particle number $N=10^{6}$ corresponds to ${\eta}{\;}{\sim}{\;}10^{-4.5}$,
which in turn implies a relaxation time 
$t_{R}{\;}{\sim}{\;}{\eta}^{-1}{\;}{\sim}{\;}10^{4.5}$.
It is, however, evident that, for a $N=10^{6}$ realisation of the thermal
equilibrium model, transitions from regularity to chaos can be important
already within a time $t<10^{2.5}$ or so! By contrast, the dynals time scale
$t_{D}{\;}{\sim}{\;}20$. 

It should also be stressed that, even if discreteness effects are too weak
to facilitate frequent transitions between regularity and chaos, they could
well play an important role in accelerating diffusion through a complex
chaotic phase space. Generic smooth potentials admitting both regular and 
chaotic orbits have chaotic phase space regions partitioned by complex 
structures associated with cantori in two dimensions and the Arnold web in 
three which, albeit not acting as absolute obstructions, serve as `entropy' 
barriers 
to slow phase space transport~\cite{LL1}. The important point, then, is 
that even very low amplitude Gaussian white noise can dramatically accelerate
diffusion through such barriers~\cite{PK}. To the extent that discreteness 
effects
can be modeled as Gaussian white noise, they too should act as a significant
source of accelerated phase space transport.

The meaning of `chaos' in the $N$-body problem is necessarily somewhat
subtle. In particular, it is important to recognise that two distinct
sources of chaos can exist, associated with physics on different scales.
Short range {\em microchaos}, associated with
close encounters between individual charges, is a generic feature of the
$N$-body problem, independent of the form of the bulk potential. However,
there is also the possibility of larger scale {\em macrochaos} which arises
if and only if, in the continuum limit, the bulk potential admits chaos. 
The important point, then, is that these two distinct sources of chaos can be 
characterised separately by different sets of Lyapunov exponents. Close
encounters trigger an exponential separation of nearby trajectories at a
rate ${\chi}_{N}$. The bulk potential triggers an exponential separation 
at a rate ${\chi}_{S}$ which is typically much smaller. 

Standard numerical computations of Lyapunov exponents yield estimates of
the much larger ${\chi}_{N}$, a quantity that does {\em not} decrease with
increasing $N$. This leads to the seemingly oxymoronic conclusion that
the $N$-body problem remains strongly chaotic for very large $N$, even if
the potential is integrable in the $N\to\infty$ limit and even if orbits
`look' nearly regular and have Fourier spectra that are nearly periodic.
The crucial point, however, is that even though microchaos remains strong
in the sense that ${\chi}_{N}$ does not decrease with increasing $N$, it
becomes progressively less important macroscopically. The scale on
which the exponential divergence saturates is comparable to a typical 
interparticle separation $r_{sep}$, a distance that decreases as $N^{-1/3}$ 
with increasing $N$. By tracking the divergence of nearby orbits, starting with
initial separations ${\ll}{\;}r_{sep}$ and continuing until the separation
becomes macroscopic, it is possible to extract estimates of both ${\chi}_{N}$
and ${\chi}_{S}$.

Finally, it should be noted that, as applied to the $N$-body problem, the 
smooth potential Lyapunov exponent ${\chi}_{S}$ does not simply quantify 
the average divergence of two nearby trajectories in a single frozen-$N$
simulation. It also quantifies the rate at which a frozen-$N$ trajectory
will diverge from a smooth potential characteristic with the same initial
condition and the rate at which orbits with the same initial condition 
diverge in different frozen-$N$ simulations, two quantities which, in some
settings, could be even more important physically.

\begin{acknowledgments}
HEK was supported in part by NSF AST-0070809. IVS and CLB were supported
in part by Department of Education Grant G1A62056.
\end{acknowledgments}

\begin{figure}
\includegraphics{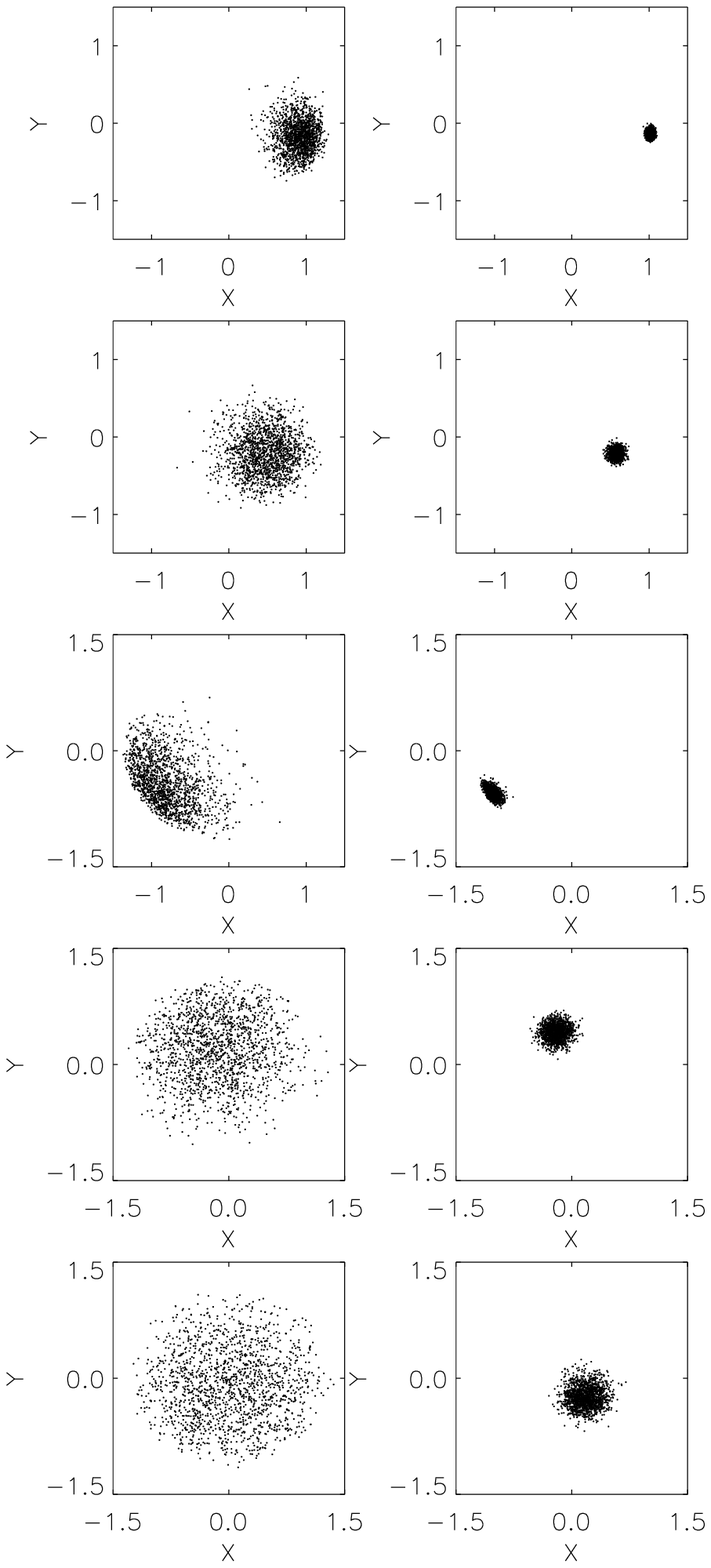}%
\caption{\label{}
The $x$ and $y$ coordinates of $1600$ initially localised points
evolved in frozen-$N$ realisations of the regular potential (4) 
with $N=10^{3.5}$
(left column) and $N=10^{5.0}$ (right column)
at different times $t$. From top to bottom,
$t=16.0$, $t=32.0$, $t=64.0$, $t=128.0$, and $t=256.0$. 
In each case, the integrations were performed with softening parameter
${e}=10^{-5}$.}
\end{figure}

\begin{figure}
\includegraphics{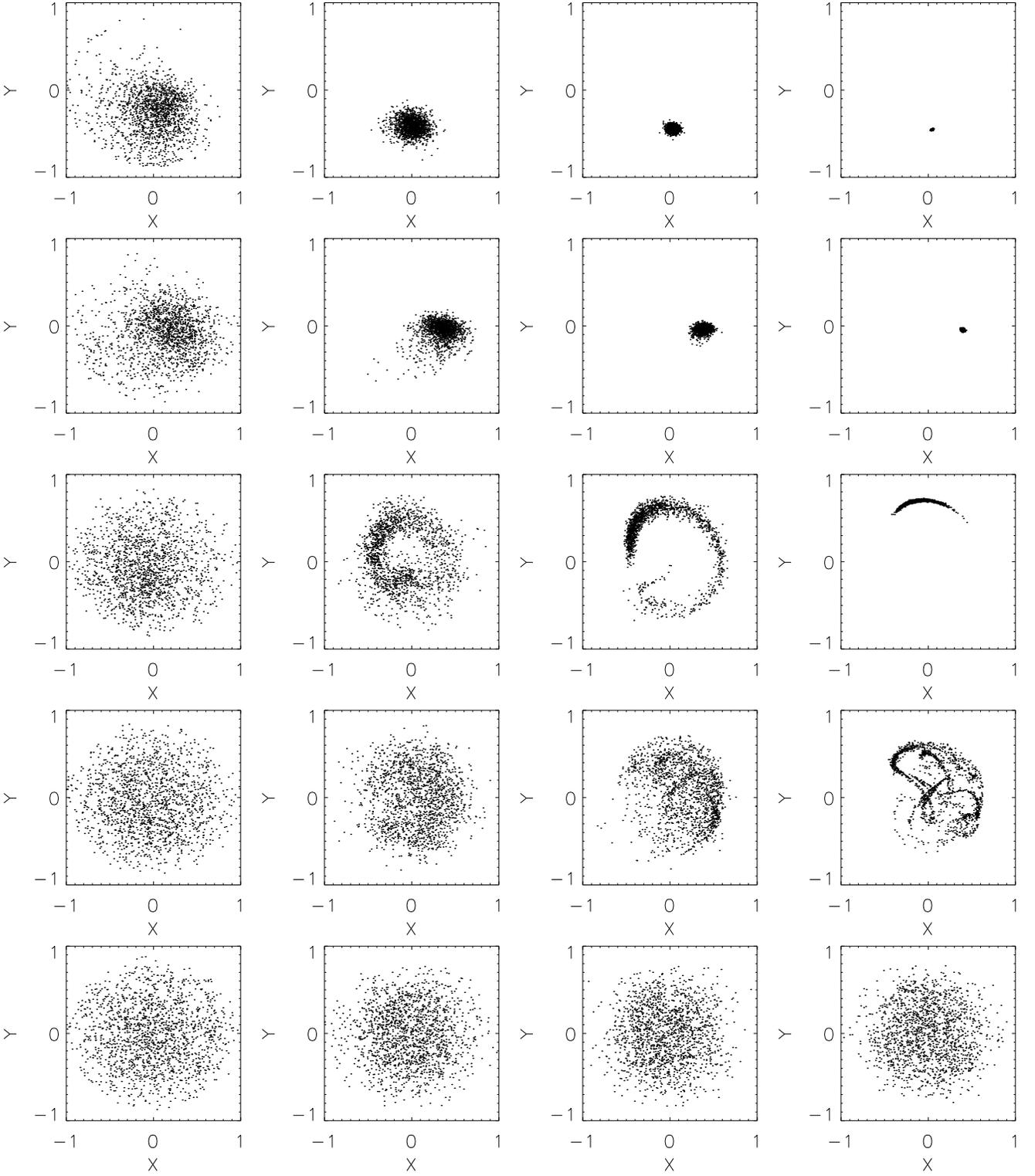}%
\caption{\label{}
The $x$ and $y$ coordinates of $1600$ initially localised points
evolved in frozen-$N$ realisations of the chaotic potential (5) with variable
$N$ at different times $t$. From left to right, one has $N=10^{3.5}$,
$N=10^{4.5}$, $N=10^{5.5}$ and the smooth potential. From top to bottom,
$t=16.0$, $t=32.0$, $t=64.0$, $t=128.0$, and $t=256.0$. Once again
${e}=10^{-5}$.}
\end{figure}

\begin{figure}
\includegraphics{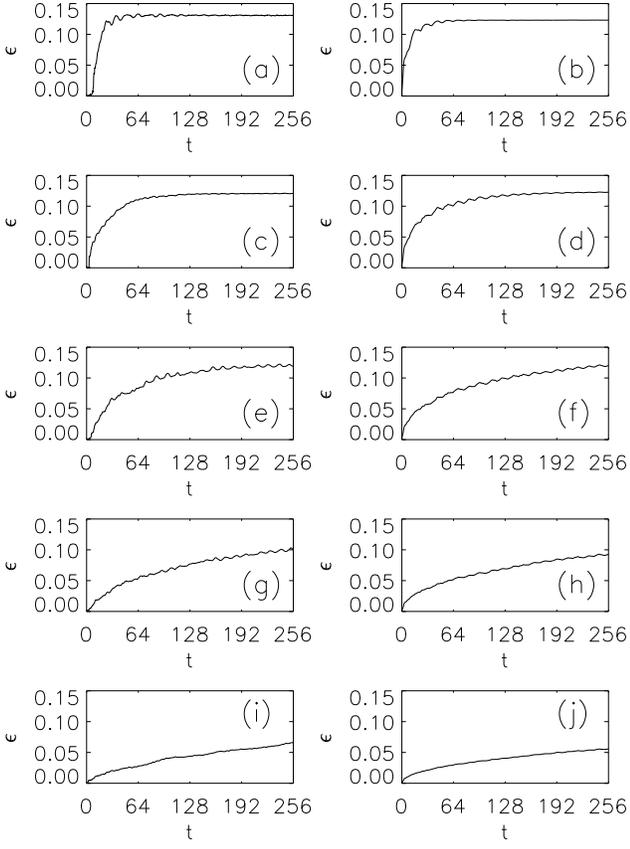}%
\caption{\label{}
The three-dimensional emittance ${\epsilon}=
({\epsilon}_{x}{\epsilon}_{y}{\epsilon}_{z})^{1/3}$ computed for the same
regular 
clump used to generate FIGURE 1, allowing for both frozen-$N$
backgrounds and energy-conserving white noise.
(a) $N=10^{3.0}$. (b) ${\eta}=10^{-2.5}$
(c) $N=10^{3.5}$. (d) ${\eta}=10^{-3.0}$
(e) $N=10^{4.0}$. (f) ${\eta}=10^{-3.5}$
(g) $N=10^{4.5}$. (h) ${\eta}=10^{-4.0}$
(i) $N=10^{5.0}$. (j) ${\eta}=10^{-4.5}$}
\end{figure}

\begin{figure}
\includegraphics{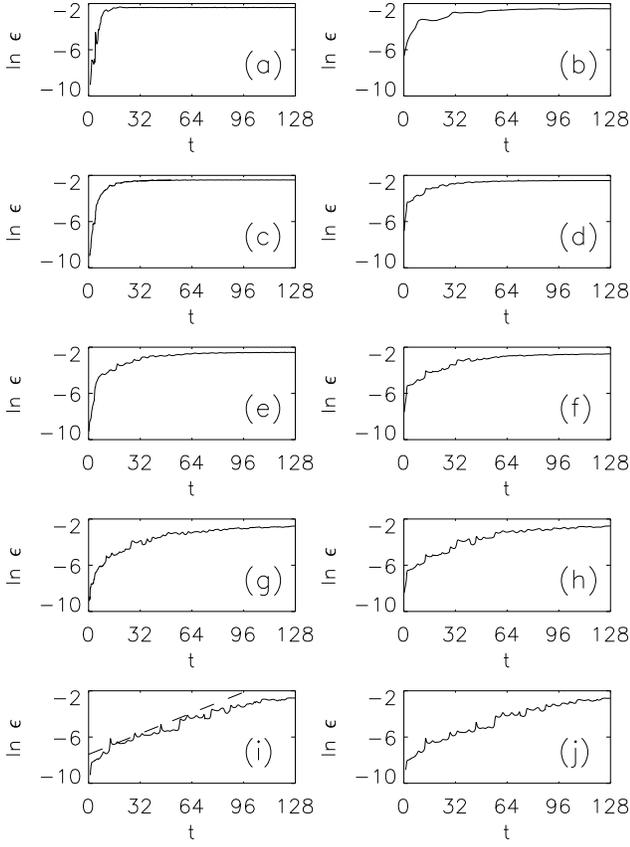}%
\caption{\label{}
The three-dimensional emittance ${\epsilon}=
({\epsilon}_{x}{\epsilon}_{y}{\epsilon}_{z})^{1/3}$ computed for the same
chaotic 
clump used to generate FIGURE 2, allowing for both frozen-$N$
backgrounds and energy-conserving white noise. Note the logarithmic scale.
(a) $N=10^{2.5}$. (b) ${\eta}=10^{-2.0}$.
(c) $N=10^{3.5}$. (d) ${\eta}=10^{-3.0}$.
(e) $N=10^{4.5}$. (f) ${\eta}=10^{-4.0}$.
(g) $N=10^{5.5}$. (h) ${\eta}=10^{-5.0}$. 
(i) The smooth potential. The dashed line corresponds to
a slope equaling the mean Lyapunov exponent ${\langle}{\chi}_{S}{\rangle}$ 
for the orbits. (j) ${\eta}=10^{-7.5}$, the strongest noise without
an appreciable effect on ${\epsilon}$.}

\end{figure}

\begin{figure}
\includegraphics{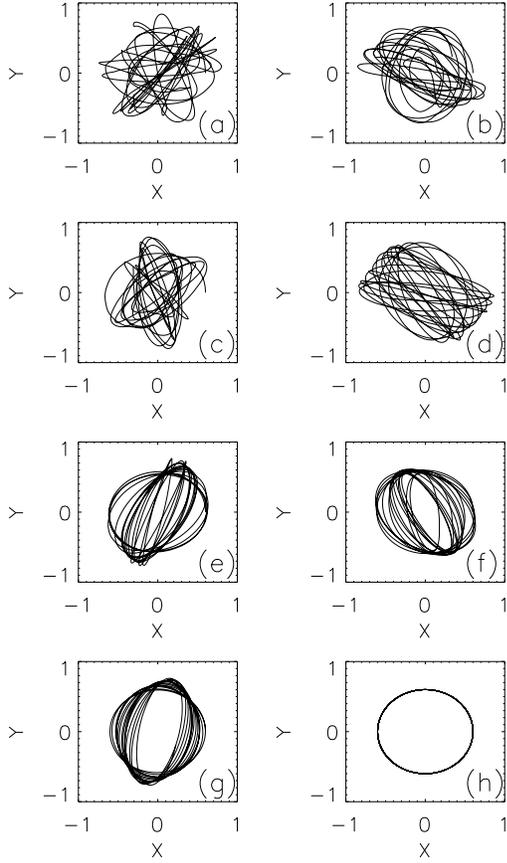}%
\caption{\label{}
The $x-y$ projection of a frozen-$N$ orbit generated from an initial
condition corresponding in the smooth potential to a circular orbit.
(a) $N=10^{2.5}$. (b) $N=10^{3.0}$
(c) $N=10^{3.5}$. (d) $N=10^{4.0}$
(e) $N=10^{4.5}$. (f) $N=10^{5.0}$
(g) $N=10^{5.5}$. (h) The smooth potential orbit. 
}
\end{figure}

\begin{figure}
\includegraphics{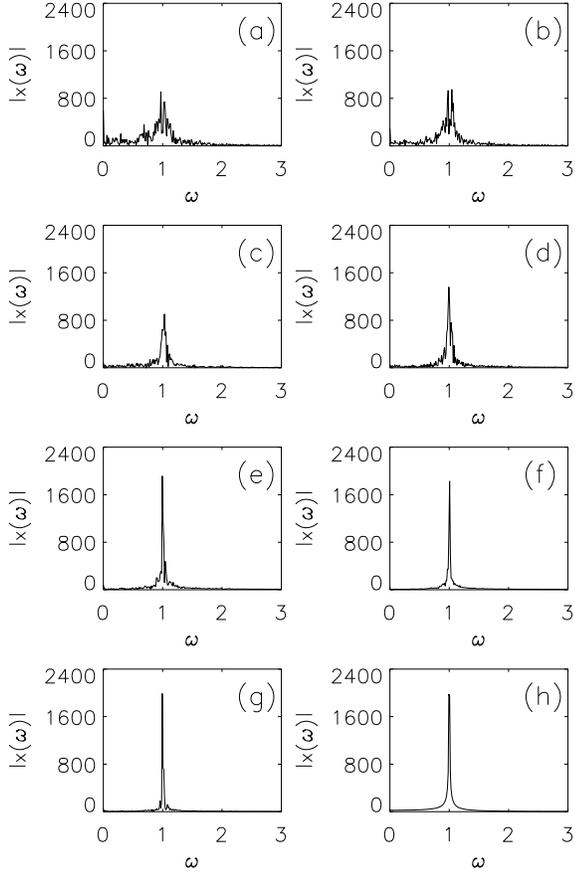}%
\caption{\label{}
The power spectrum $|x({\omega})|$ for the orbits exhibited in the
preceding Figure.
(a) $N=10^{2.5}$. (b) $N=10^{3.0}$
(c) $N=10^{3.5}$. (d) $N=10^{4.0}$
(e) $N=10^{4.5}$. (f) $N=10^{5.0}$
(g) $N=10^{5.5}$. (h) The smooth potential orbit.}
\end{figure}

\begin{figure}
\includegraphics{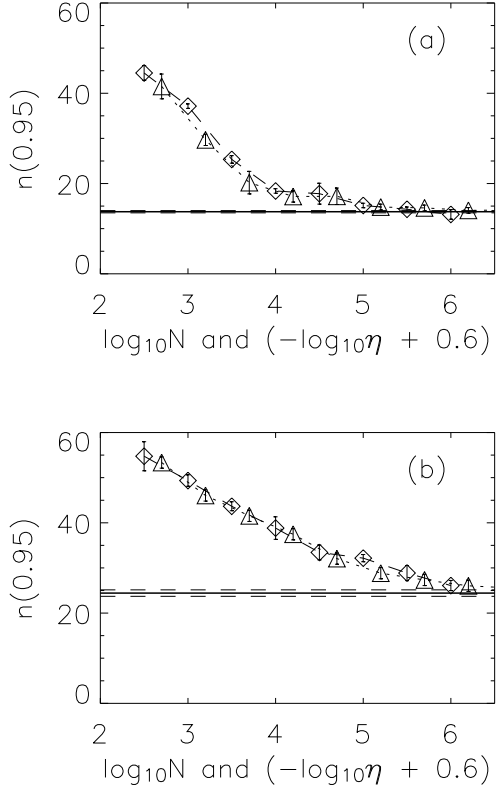}%
\caption{\label{} (a) Diamonds show the complexity $n(0.95)$, defined as the 
mean number of frequencies required to capture 95\% of the total power, 
computed for a collection of 100 initial conditions integrated in frozen-$N$ 
realisations of the integrable Model 1 with variable $N$. Triangles show the 
same quantity for the same initial conditions integrated in the smooth 
potential but subjected to Gaussian white noise with variable ${\eta}$ 
The solid line exhibits the mean complexity for orbits evolved in the
unperturbed smooth potential. (b) The 
same as (a), generated from the same initial conditions but now computed
for the strongly chaotic Model 2.
}
\end{figure}



\begin{figure}
\includegraphics{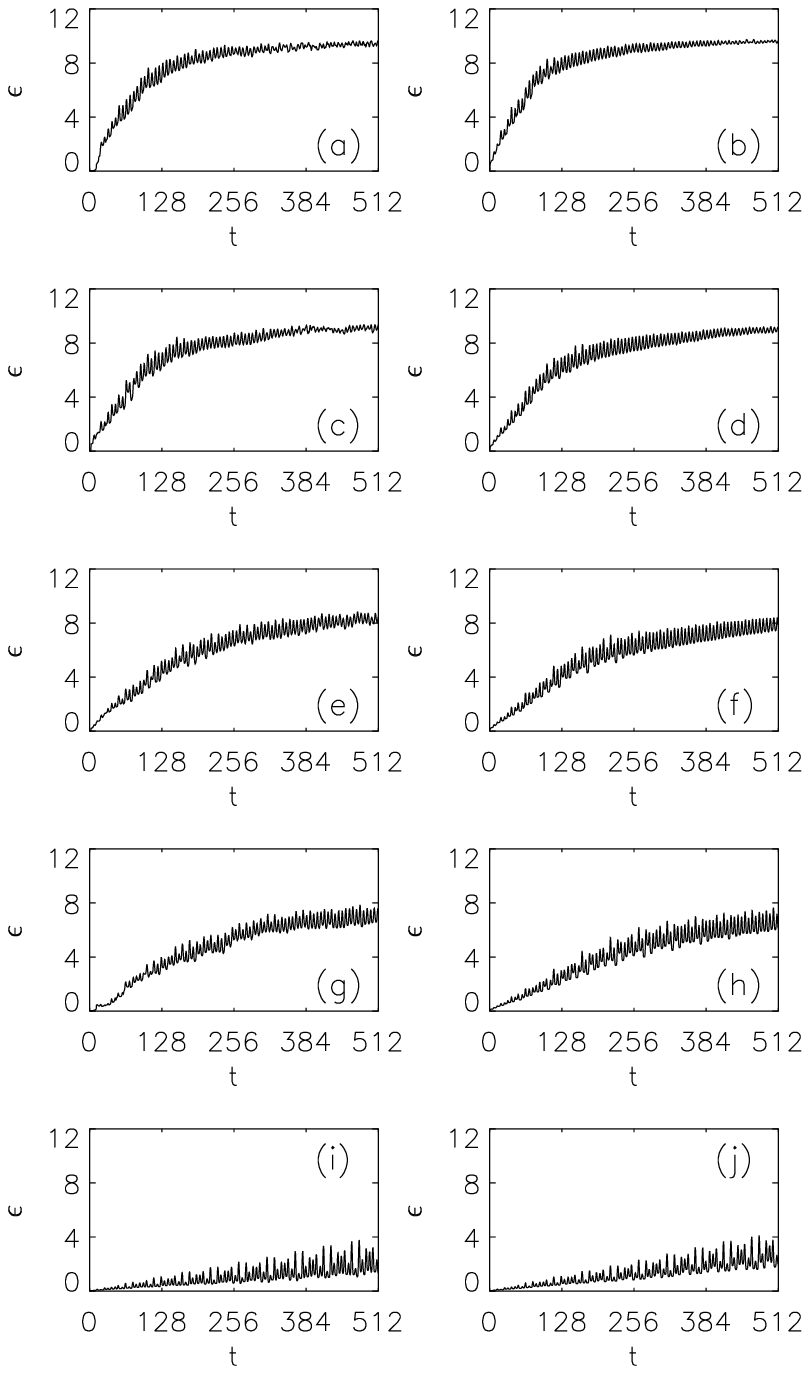}%
\caption{\label{} The three-dimensional emittance ${\epsilon}$ computed
for 
a clump of regular initial conditions for the thermal equilibrium
model, allowing for both frozen-$N$ backgrounds and energy conserving white
noise. (a) $N=10^{4.5}$. (b) ${\eta}=10^{-3.0}$.
(c) $N=10^{5.0}$. (d) ${\eta}=10^{-3.5}$. (e) ${\eta}=10^{5.5}$. 
(f) ${\eta}=10^{-4.0}$. (g) $N=10^{6.0}$. (h) ${\eta}=10^{-4.5}$.
(i) Unperturbed motion in the smooth potential.
(j) ${\eta}=10^{-6.5}$, the largest value of ${\eta}$ that does not 
significantly impact emittance growth.}
\end{figure}

\begin{figure}
\includegraphics{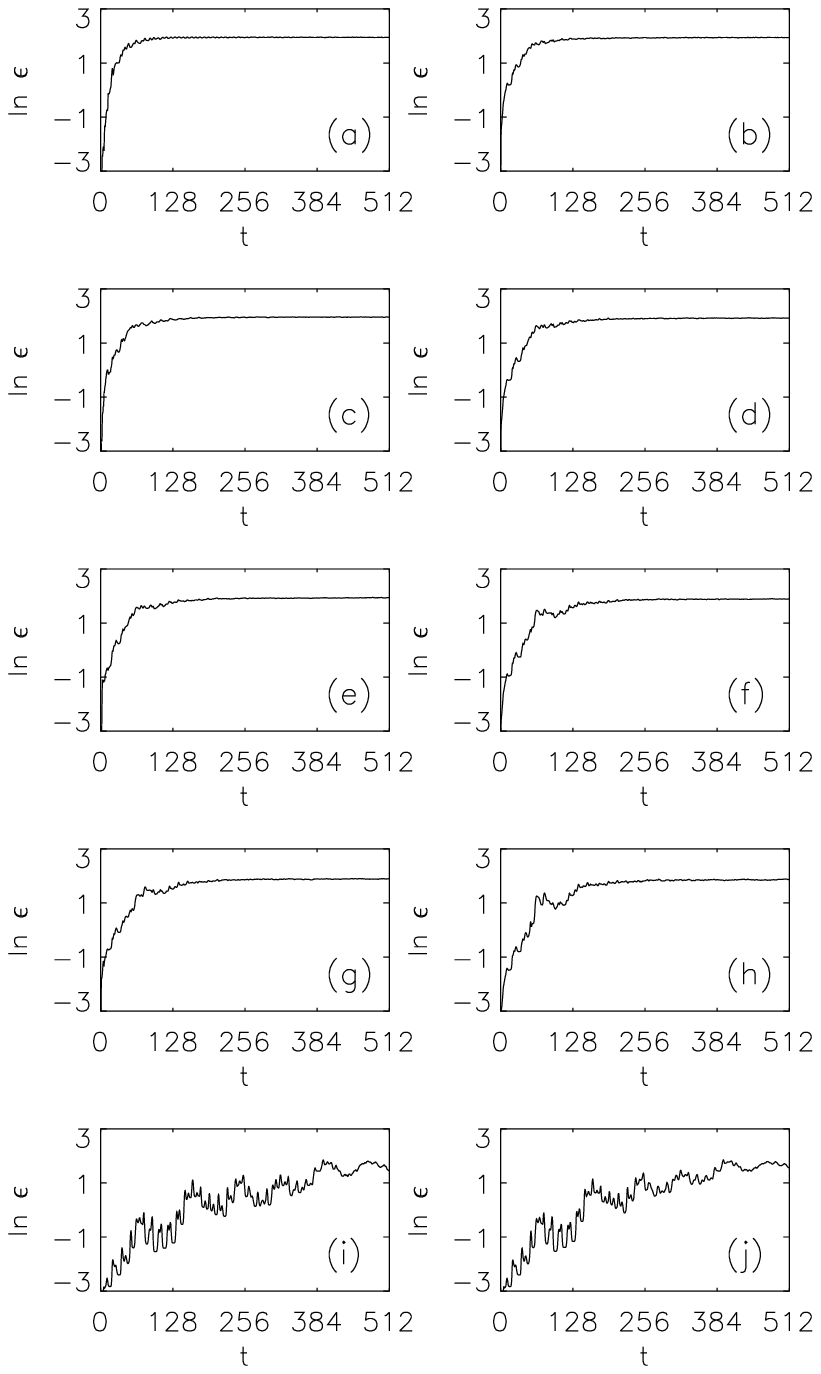}%
\caption{\label{}  The three-dimensional emittance ${\epsilon}$ computed
for 
a clump of chaotic initial conditions for the thermal equilibrium
model, again allowing for both frozen-$N$ backgrounds and energy conserving 
white noise. (a) $N=10^{4.5}$. (b) ${\eta}=10^{-3.0}$.
(c) $N=10^{5.0}$. (d) ${\eta}=10^{-3.5}$. (e) $N=10^{5.5}$. 
(f) ${\eta}=10^{-4.0}$. (g) $N=10^{6.0}$. (h) ${\eta}=10^{-4.5}$.
(i) Unperturbed motion in the smooth potential.
(j) ${\eta}=10^{-8.0}$, the largest value of ${\eta}$ that does not 
significantly impact emittance growth.}
\end{figure}

\begin{figure}
\includegraphics{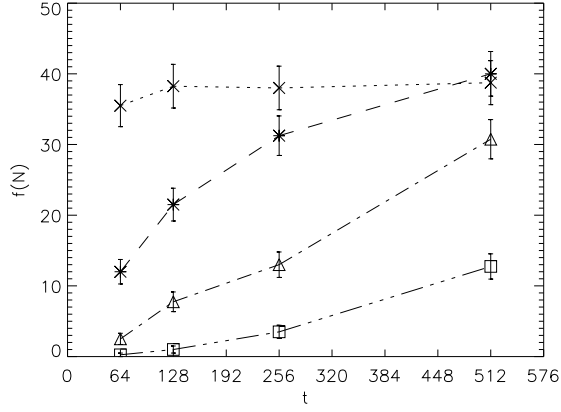}%
\caption{\label{} The percentage of frozen-$N$ orbits generated from 
a clump of regular initial conditions and evolved in the thermal equilibrium
model which, at time $t$, have been converted to chaotic orbits. From top to 
bottom, the curves correspond to frozen-$N$ backgrounds with $N=10^{4.5}$, 
$N=10^{5.0}$, and $N=10^{5.5}$, and $N=10^{6.0}$.
}
\end{figure}

%
%
%

\begin{figure}
\includegraphics{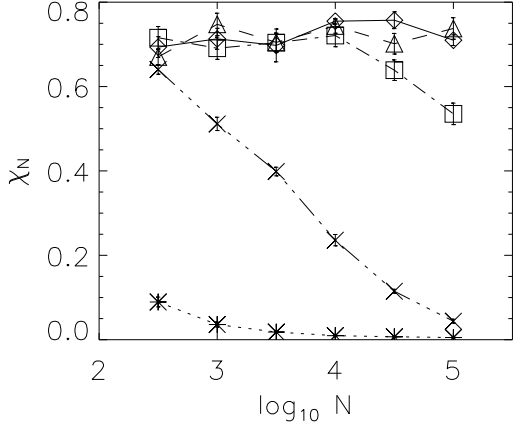}%
\caption{\label{}
Mean value of the largest Lyapunov exponent ${\chi}_{N}$ as a function of
$N$ for different choices of softening parameter ${e}$, computed for
regular initial conditions evolved in Model 1. 
${e}=10^{-5}$: solid line and diamonds. 
${e}=10^{-4}$: dashed line and squares. 
${e}=10^{-3}$: dot-dashed line and triangles. 
${e}=10^{-2}$: triple-dot-dashed line and crosses.
${e}=10^{-1}$: dots and stars.}
\end{figure}

\begin{figure}
\includegraphics{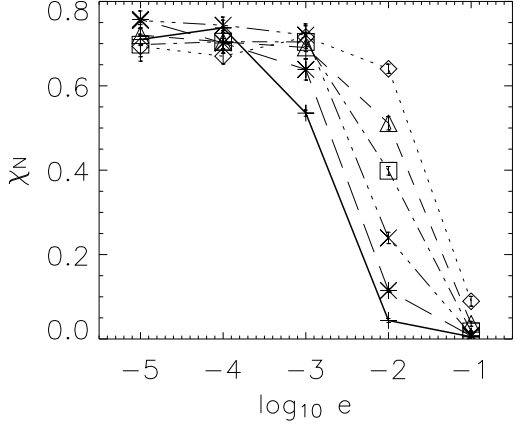}%
\caption{\label{}
Mean value of the largest Lyapunov exponent ${\chi}_{N}$ as a function of
softening parameter ${e}$ for different choices of $N$, computed for
regular initial conditions evolved in Model 1. 
$N=10^{5.0}$: solid line and pluses. 
$N=10^{4.5}$: broad dashed line and stars. 
$N=10^{4.0}$: triple-dot-dashed line and crosses. 
$N=10^{3.5}$: dot-dashed line and squares.
$N=10^{3.0}$: dashed line and triangles.
$N=10^{2.5}$: dots and diamonds.}
\end{figure}

\begin{figure}
\includegraphics{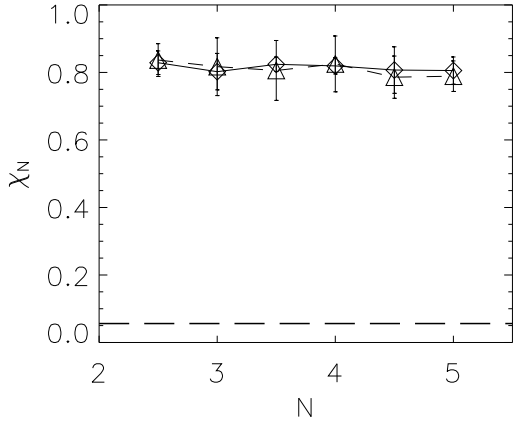}%
\caption{\label{}
Mean value of the largest Lyapunov exponent ${\chi}_{N}$ as a function of
$N$ for the same initial condition integrated in the integrable Model 1
(solid curve) and the chaotic Model 2  (short dashed curve). In both cases,
${e}=10^{-5}$. The broad-dashed line at the bottom corresponds to
the smooth potential Lyapunov exponent ${\chi}_{S}$ for the same initial
condition integrated in the smooth potential corresponding to Model 2.}
\end{figure}

\begin{figure}
\includegraphics{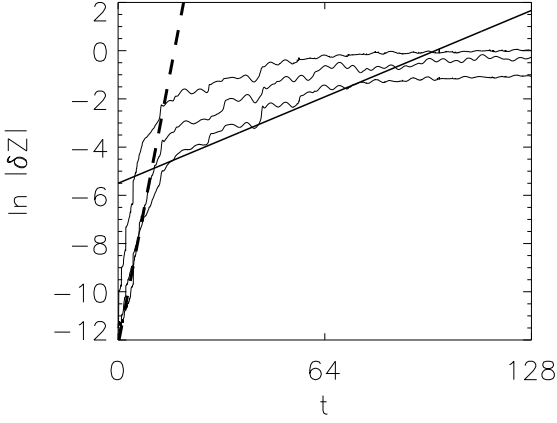}%
\caption{\label{}
The mean phase space separation $|{\delta}Z(t)|$ for 50 nearby pairs of initial
conditions evolved in frozen-$N$ realisations of the chaotic Model 2 with 
(from top to bottom)
$N=10^{5.0}$, $10^{5.5}$, and $10^{6.0}$. The solid line has a slope 
${\chi}_{S}=0.056$, equal to the the mean value of the largest smooth
potential Lyapunov exponent. The dashed line has a slope ${\chi}_{N}=0.82$,
equal to the mean value of the largest $N$-body exponent.}
\end{figure}

\begin{figure}
\includegraphics{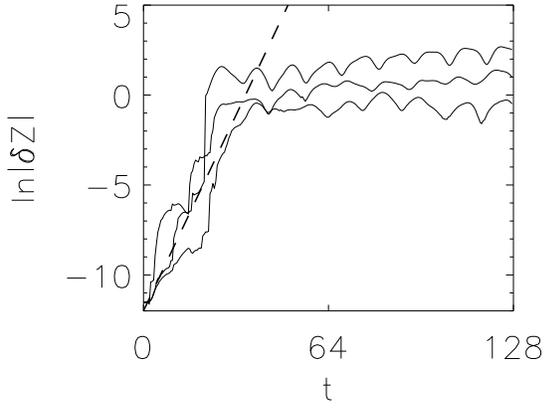}%
\caption{\label{}The same as the preceding for regular orbits in the thermal
equilibrium model. The dashed line again has a slope equal to the mean value
of the largest $N$-body Lyapunov exponent. Note the absence of the second 
exponential phase.}
\end{figure}


\begin{figure}
\includegraphics{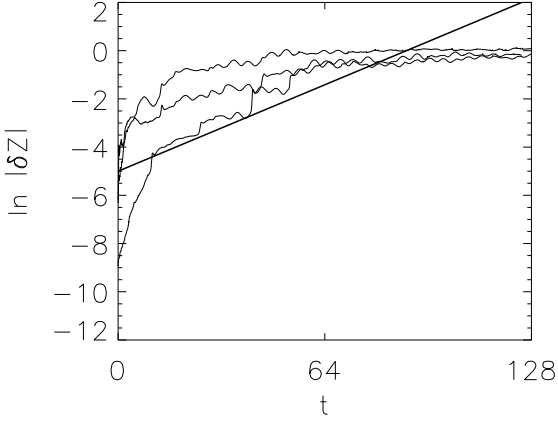}%
\caption{\label{}
The mean phase space separation $|{\delta}Z(t)|$ for 50 initial conditions
evolved in two different frozen-$N$ realisations of Model 2 with 
(from top to bottom)
$N=10^{5.0}$, $10^{5.5}$, and $10^{6.0}$. The solid line again has a slope 
${\chi}_{S}=0.056$, equal to the mean value of the largest smooth
potential Lyapunov exponent.}
\end{figure}

\begin{figure}
\includegraphics{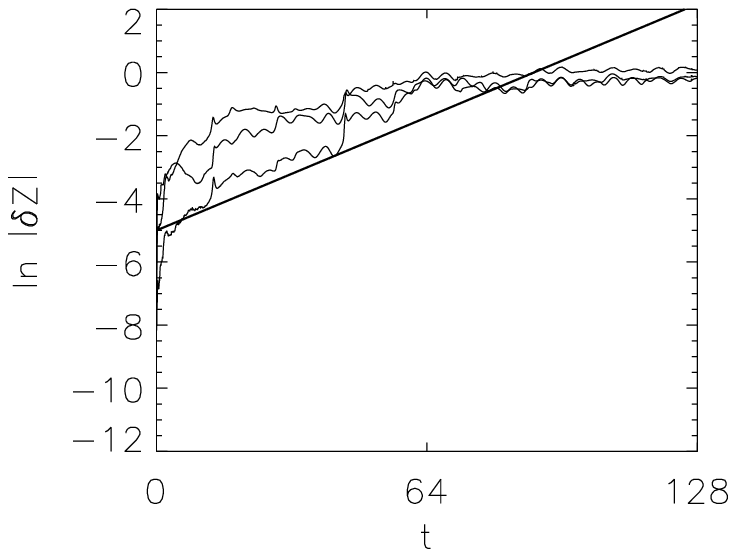}%
\caption{\label{}
The mean phase space separation $|{\delta}Z(t)|$ for 50 initial conditions
evolved in Model 2, both in the smooth potential and in a frozen-$N$ system
with (from top to bottom)
$N=10^{5.0}$, $10^{5.5}$, and $10^{6.0}$. The solid line again has a slope 
${\chi}_{S}=0.056$, equal to the mean value of the largest smooth
potential Lyapunov exponent.}
\end{figure}


\begin{references}
\bibitem{Foot} Recent numerical simulations of interacting 
self-gravitating systems performed by Hemsendorff and Merritt~\cite{HM} are
completely consistent with this prediction, the data there suggesting that
${\chi}_{N}(N)$ should increase logarithmically as $N$ increases.
\bibitem{smo} By increasing the value of the softening parameter ${\epsilon}$,
one can `force' the frozen-$N$ system to more closely resemble the smooth
density distribution; but this comes at the cost of supressing physical effects
associated with finite $N$ which are present in real systems.
\end{references}
\end{document}